\documentclass[nofootinbib,twocolumn,showpacs,preprintnumbers,pre,aps]{revtex4-1}
\usepackage{graphicx}% Include figure files
\usepackage{bm}% bold math
\usepackage{amsmath,color}
\usepackage{amssymb}
\usepackage{lipsum}
\DeclareMathOperator{\Tr}{Tr}

\begin{document}
%\draft
%\bibliographystyle{prsty}

\title{Relaxation dynamics of two-component fluid bilayer membranes}

\author{Ryuichi Okamoto$^{1,2}$}

\author{Yuichi Kanemori$^{1}$}

\author{Shigeyuki Komura$^{1,2,3}$}

\author{Jean-Baptiste Fournier$^{2}$}

\affiliation{
$^{1}$Department of Chemistry, Graduate School of Science and Engineering,
Tokyo Metropolitan University, Tokyo 192-0397, Japan\\
$^{2}$Universit\'e Paris Diderot, Sorbonne Paris Cit\'e, 
Laboratoire Mati\`ere et Syst\`emes Complexes (MSC), 
UMR 7057 CNRS, F-75205 Paris, France\\
$^{3}$Kavli Institute for Theoretical Physics China, CAS,
Beijing 100190, China}

\date{\today}
%\date{version 8 by RO/SK, November 2, 2015}

\begin{abstract} 
We theoretically investigate the relaxation dynamics of a nearly-flat binary lipid bilayer 
membrane by taking into account the membrane tension, hydrodynamics of the surrounding fluid, 
inter-monolayer friction and mutual diffusion. Mutual diffusion is the collective irreversible process that leads to homogenization of the density difference between the two lipid species.
We find that two relaxation modes associated with the mutual diffusion appear in addition 
to the three previously discussed relaxation modes reflecting the bending and compression
of the membrane. 
Because of the symmetry, only one of the two diffusive mode is coupled to the bending mode. 
The two diffusive modes are much slower than the bending and compression modes in the 
entire realistic wave number range.
This means that the long time relaxation behavior is dominated by the mutual diffusion in 
binary membranes. 
The two diffusive modes become even slower in the vicinity of the unstable region 
towards phase separation, while the other modes are almost unchanged. 
In short time scales, on the other hand, the lipid composition heterogeneity induces in-plane 
compression and bending of the bilayer. 
\end{abstract}
%\pacs{67.25.dm, 87.16.D-, 82.70.Uv, 87.16.dj}

\maketitle

%%%%%%%%%%%%%%%%%%%%%%
\section{Introduction}
%%%%%%%%%%%%%%%%%%%%%%
\label{introduction}

Much attention has been paid to artificial lipid bilayer membranes as model systems 
of biological cell membranes~\cite{AlbertsBook}. 
They exhibit a wide variety of complex phenomena in both statics and dynamics, since 
lipid densities, membrane deformation and surrounding fluids are coupled to each 
other~\cite{Lipowsky95}. 
Dynamical properties of lipid membranes near the equilibrium is characterized by 
wavenumber dependent relaxation rates. 
In the early theoretically studies, the relaxation rate of a single-component membrane 
was discussed by regarding a membrane as an elastic sheet having out-of-plane deformation, 
and further surrounded by a three-dimensional (3D) fluid. 
Neglecting the bilayer structure, several authors predicted that the relaxation of 
the bending mode is dominated by the bending rigidity and the viscosity of the surrounding 
bulk fluid~\cite{Kramer,Brochard}.

Later, Seifert and Langer considered the inter-monolayer friction and the two-dimensional 
(2D) hydrodynamics of each monolayer, and obtained another relaxation mode associated with 
the density difference between the two monolayers~\cite{Seifert}.
They found that the relaxation of the density fluctuation is dominated by the inter-monolayer 
friction and is relevant to the slow dynamics characterized by large wave numbers, whereas 
the relaxation of the bending mode is relevant for small wave numbers if the membrane 
surface tension is not acting.
A somewhat similar theory was also developed in ref.~\cite{Yeung}.
The predicted mode crossing behavior has been supported by several 
experiments~\cite{Pfeiffer,Pott} and by molecular dynamics simulations~\cite{Shkulipa}. 
More recently, some experiments reported a chemically induced tubule growing from a giant 
unilamellar vesicle (GUV)~\cite{JBprl,Bitbol1}.
In these studies, they showed that the interplay between the faster bending relaxation 
and the slower density relaxation on the scale of tens of micrometers plays 
an essential role.

In recent years, both the statics and dynamics of multi-component lipid membranes have been 
extensively studied because 2D phase separation takes place in a certain range of 
temperature and composition~\cite{VK05,HVK09,SK_DA_Review}.
Studies on the dynamics of multi-component membranes can be classified into two
categories; (i) dynamics of lateral phase separation below the phase separation 
temperature, and (ii) dynamics of concentration fluctuations above the critical 
temperature. 
For the details of the domain growth dynamics in the lower temperatures, which 
is not the subject of the present work, readers are referred to ref.~\cite{SK_DA_Review}.
Experimentally, Honerkamp-Smith \textit{et al.} have investigated the dynamics 
of concentration fluctuations in ternary GUVs and showed that the dynamic critical 
exponent crosses over from a 2D value to a 3D one as the critical temperature is
approached from above~\cite{KellerDynamics}. 
The dynamics of concentration fluctuations in membranes was first modeled by 
Seki \textit{et al.}~\cite{SKI07} and later extended by Inaura and 
Fujitani~\cite{Inaura}.
Ramachandran \textit{et al.}  used the general mobility tensor to numerically 
calculate the effective diffusion coefficient of concentration 
fluctuations~\cite{RKSI11}.
In these theoretical works, however, they did not take into account the 
out-of-plane membrane deformation nor the membrane bilayer structure.

In biomembranes of living cells, the two monolayers have in general different 
compositions, with a unique asymmetry between the inner and outer leaflets.
Furthermore, the two leaflets are not independent, but rather interact strongly 
with each other due to various physical and chemical 
mechanisms~\cite{May09}.
Some experiments have shown strong positional correlation and domain registration 
between domains across the two membrane leaflets~\cite{Collins08,CK08}, while 
some papers reported the anti-registration of domains in different 
leaflets~\cite{Regen1,Regen2,Longo}.
Inspired by the experiments, several simulations have been 
performed~\cite{Stevens05,Sachs11} and some phenomenological models have been 
proposed~\cite{WLM07,Schick08} to describe the phase separation in such coupled 
leaflets.
Hirose \textit{et al.} considered a coupled bilayer composed of two modulated 
monolayers and discussed the static and dynamic properties of concentration 
fluctuations above the transition temperature~\cite{HKA09,HKA12}.

The purpose of this paper is to investigate the relaxation dynamics of a 
binary lipid bilayer membrane in one phase region (rather than phase separated
membranes in two phase state).
In such membranes, the interplay of various important effects, such as inter-monolayer 
friction and composition-deformation coupling, leads to a complex behavior. 
 In particular, as in usual 3D multi-component fluids, a chemical potential gradient, and thus mutual diffusion, are induced by the inhomogeneity of the density difference between the two lipid species. Such mutual diffusion leads to homogenization of the density difference in each monolayer, as an irreversible process.
In this paper, we take into account the membrane surface tension, 2D hydrodynamics of 
each monolayer, mutual diffusion in each monolayer, 3D hydrodynamics of the surrounding fluid, 
and inter-monolayer friction, whereas the flip-flop motion of the lipid molecules between the
two leaflets is not included. 
In our model, the sources of the energy dissipation are the viscosities of the monolayers 
and the surrounding fluid, the inter-monolayer friction, and the mutual diffusion in each 
monolayer. 
We find that the two relaxation modes associated with the mutual diffusion appear in addition 
to the three previously discussed relaxation modes~\cite{Seifert}.
These two diffusive modes turn out to be much slower than the other hydrodynamic modes,
and become even slower in the vicinity of the unstable region towards the phase separation.

This paper is organized as follows. 
In sect.~\ref{freeenegy}, we present the free energy functional of a binary lipid bilayer 
membrane by assuming that the membrane deformation and the density deviations from the 
respective average values are small. 
The whole set of dynamic equations are introduced in sect.~\ref{dynamics}, and the surrounding 
flow field is integrated out to obtain the relaxation equations for the membrane variables.
In sect.~\ref{results}, thermodynamic stability of the one phase state and the wave 
number dependencies of various relaxation modes are discussed in detail for both 
small and moderate surface tension cases.
We also present our numerical study of the domain relaxation dynamics.
Finally, sect.~\ref{summary} is devoted for summary and discussion.

%%%%%%%%%%%%%%%%%%%%% 
\section{Free Energy}
%%%%%%%%%%%%%%%%%%%%%
\begin{figure}
\label{freeenegy}

\includegraphics[scale=0.43]{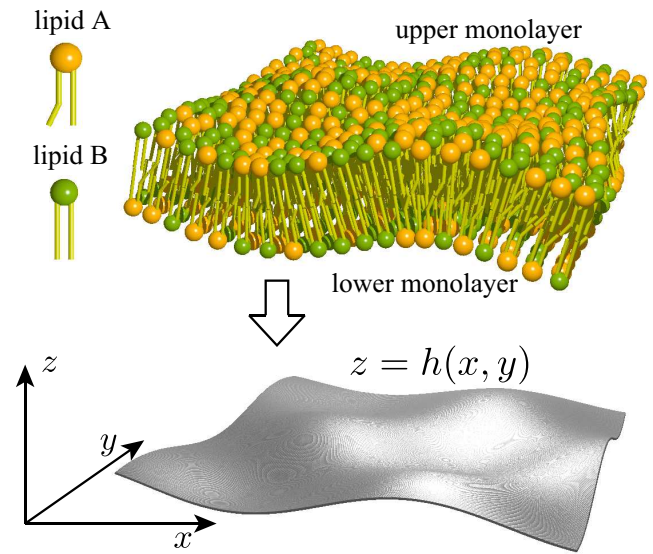}
\caption{Schematic representation of a two-component fluid bilayer membrane consisting 
of lipid A and lipid B.
The hydrophobic chains are arranged in a back-to-back configuration to form a bilayer. 
The surface at which the upper and lower monolayers are in contact with each other is 
defined as the mid-surface.
The membrane shape is expressed by the height $z=h(x,y)$ of the mid-surface measured 
from the $z=0$ plane.}
%\label{fig.1}
\label{Figillustration}
\end{figure}

A binary lipid bilayer membrane consists of lipid A and lipid B
as schematically presented in fig.~\ref{Figillustration}. 
In the presence of the surrounding fluid, the hydrophobic tails of lipid molecules face 
each other to form a bilayer structure, while the hydrophilic heads are in contact 
with the outer fluid. 
The surface at which the hydrocarbon tails are in contact with each other is defined 
as the mid-surface. 
Using the height $h(x,y)$ of the mid-surface from the $z=0$ plane in the 3D Euclidean space,
we express the shape of a nearly flat membrane using the Monge gauge, {\it i.e.}, $z=h(x,y)$.

Let us write $\psi _{\rm J}^\pm$ the areal mass densities of lipid J 
(${\rm J} = {\rm A,B}$) in the upper ($+$) and lower ($-$) monolayers, respectively.  
The total free energy of the bilayer membrane is generally given by the form
\begin{equation}
F=\int {\rm d}^2x \, 
\sqrt{g} \, f_{\rm tot}(H, \psi_{\rm J}^\pm, \nabla_\perp\psi_{\rm J}^\pm), 
\label{Ftot}
\end{equation}
where $\nabla_\perp$ is the gradient operator along the membrane surface, 
$g$ the determinant of the metric tensor, 
and $\int {\rm d}^2x$ denotes the integration with respect to $x$ and $y$. 
Within the lowest order in $h$, we have $\nabla_\perp\simeq \tilde\nabla$ where 
$\tilde\nabla=(\partial_x,\partial_y)$ is the 2D gradient operator in the 
projected plane.
The areal free energy density $f_{\rm tot}$ depends on the mean curvature $H$, the densities 
and their spacial derivatives.  
In general, the free energy also depends on the temperature, but we shall not write the 
temperature dependence of any quantities explicitly. 
For small membrane deformations ($|\tilde\nabla h| \ll 1$), $H$ and $g$ are approximated 
as $H \simeq (\tilde{\nabla}^2 h)/2$ and $g\simeq1+(\tilde{\nabla} h)^2$, respectively,
within the lowest order in $h$.

We assume in this paper that the upper and the lower monolayers have the same number of 
lipid molecules, namely, 
$\int {\rm d}^2x \, \sqrt{g} \, \psi_{\rm J}^+= \int {\rm d}^2x \, \sqrt{g} \, \psi_{\rm J}^-$.
We introduce the reference mass densities of the lipid molecules $\psi_{\rm J0}$ as the 
spacial average of the densities for a flat membrane (or projected mass densities). 
Then the conservation law for the lipid molecules is written as
\begin{equation}
\int {\rm d}^2x \, \sqrt{g} \, \psi_{\rm J}^\pm = \psi _{\rm J0} \int {\rm d}^2x. 
\label{conservation1}
\end{equation}
We further define the normalized density deviations as
\begin{equation}
\rho_{\rm J}^\pm = \frac{\psi_{\rm J}^\pm}{\psi_{\rm J0}}-1.  
\label{rhoJ}
\end{equation}
With the aid of eq.~(\ref{rhoJ}), the conservation law eq.~(\ref{conservation1}) can be 
rewritten as
\begin{equation}
\int {\rm d}^2x\, \sqrt{g}\, \rho_{\rm J}^\pm \simeq
- \frac{1}{2} \int {\rm d}^2x \,  (\tilde\nabla h)^2, 
\label{conservation2}
\end{equation}
up to the second order in $h$. 
Notice that the integral in the left hand side does not vanish exactly because 
$\psi_{\rm J0}$ is the projected average density.

%%%%%%%%%%%%%%%%%%%%%%%%%%%%%%%%%
\subsection{Bilinear free energy}

Hereafter we assume that the membrane is weakly deformed and the density deviations are 
small enough so that $h$ and $\rho_{\rm J}^\pm$ can be treated as small variables. 
Then $\sqrt{g}$ and $f_{\rm tot}$ in eq.~(\ref{Ftot}) can be expanded about the 
reference state ($h=0$, $\psi_{\rm J}^\pm=\psi_{\rm J0}$)
with respect to the small variables 
$h$, $\rho_{\rm J}^\pm$, and $\tilde\nabla\rho_{\rm J}^\pm$. 
The total free energy is given by the sum of three contributions
\begin{equation}
F=F_{\rm def}+F_{\rm coup}+F_{\rm grad}, 
\end{equation}
where $F_{\rm def}$ is the deformation part, 
$F_{\rm coup}$ the coupling part, and $F_{\rm grad}$ the gradient part.
Each part will be explained in order.

First the deformation part $F_{\rm def}$ is given by
\begin{equation}
F_{\rm def}=\int {\rm d}^2x 
\left[ \frac{\sigma}{2} (\tilde{\nabla} h)^2+
\frac{\kappa}{2}(\tilde{\nabla}^2 h)^2 \right], 
\label{Fdef}
\end{equation}
where $\sigma$ is the membrane surface tension and $\kappa$ the bending rigidity.
 The surface tension $\sigma$ is expressed in terms of $f_{\rm tot}$ in eq.~(\ref{Ftot}) as 
\begin{equation}
\sigma=f_{\rm tot}-\sum_{\epsilon=+,-}\sum_{{\rm J}={\rm A, B}} 
\frac{\partial f_{\rm tot}}{\partial \psi_{\rm J}^\epsilon} \psi_{\rm J0}, 
\label{sigma}
\end{equation} 
where $f_{\rm tot}$ and its derivatives $\partial f_{\rm tot}/\partial \psi_{\rm J}^\epsilon$ 
are evaluated at the reference state. 
In deriving eqs.~(\ref{Fdef}) and (\ref{sigma}), we have made use of 
$\sqrt{g}\simeq 1+(\tilde\nabla h)^2/2$, 
$\partial f_{\rm tot}/\partial \rho_{\rm J}^\pm=
(\partial f_{\rm tot}/\partial \psi_{\rm J}^\pm)\psi_{\rm J0}$ and eq.~(\ref{conservation2}). 
The right hand side of eq.~(\ref{sigma}) can be identified as the (negative) in-plane pressure 
for a flat membrane~\cite{Pressure}.

The coupling part $F_{\rm coup}$ consists of all the possible bilinear couplings between 
$H$ and $\rho^\pm_{\rm J}$. 
For later convenience, we introduce the normalized total mass density deviation
\begin{equation}
\rho^\pm= \frac{\sum_{\rm J}\psi_{\rm J}^\pm}{\sum_{\rm J}\psi_{\rm J0}}-1=
\frac{\sum_{\rm J}\psi_{\rm J0}\rho_{\rm J}^\pm}{\sum_{\rm J}\psi_{\rm J0}}, 
\label{rho}
\end{equation}
and the normalized mass density difference 
\begin{equation}
\phi^\pm= \rho_{\rm A}^\pm-\rho_{\rm B}^\pm. 
\label{phi}
\end{equation}
We express $F_{\rm coup}$ in terms of bilinear couplings between
$H \simeq (\tilde{\nabla}^2 h)/2$, $\rho^\pm$ and $\phi^\pm$ rather than 
those between $H$ and $\rho^\pm_{\rm J}$. 
With this choice of variables, the dynamic equations will be simplified as we will 
show in the next section.
Since we have five independent variables, there should be in principle fourteen coupling 
parameters in $F_{\rm coup}$~\cite{coupling}.
However, we can reduce the number of coupling parameters by using the invariance of the 
system under the interchange of the upper and the lower monolayers. 
For instance, the coupling parameter for $(\rho^+)^2$ should be the same for $(\rho^-)^2$. 
Also the coupling parameter for $\rho^+(\tilde\nabla^2h)$
should have the same magnitude but with an opposite sign of that for 
$\rho^-(\tilde\nabla^2h)$.
Using these symmetric properties, we are left with eight coupling parameters. 
Furthermore, it is convenient to absorb two of them, $d$ and 
(dimensionless) $\nu$, in the following redefinitions of the variables:
\begin{equation}
\alpha^\pm\equiv \rho^\pm \pm d (\tilde\nabla^2h), \quad 
\beta^\pm \equiv \phi^\pm \pm \nu d (\tilde\nabla^2h). 
\label{alphabeta}
\end{equation}
The two lengths $d$ and $\nu d$ can be interpreted as the distances between the 
membrane mid-surface and the two effective neutral surfaces~\cite{Seifert}.
Introducing the parameters $k$ and $\Lambda_i$ $(i=1,\cdots, 5)$, we can write 
$F_{\rm coup}$ in the form
\begin{align}
F_{\rm coup}   = &\frac{k}{2}\int {\rm d}^2x  
\Bigg[ \sum_{\epsilon =+,-} \left\{ (\alpha^\epsilon)^2+\Lambda_1(\beta^\epsilon)^2
+\Lambda_2 \alpha^\epsilon\beta^\epsilon \right\} \nonumber \\
& +\Lambda_3 \alpha^+\alpha^- + \Lambda_4 \beta^+\beta^- + 
\Lambda_5( \alpha^+\beta^-+\alpha^-\beta^+) \Bigg]. 
\label{Fcoup}
\end{align}
Here $k$ has the dimension of areal compression modulus, and $\Lambda_i$ are 
the dimensionless parameters of order unity.

Within the lowest order in the membrane deformations and density deviations, we can 
approximate as $\nabla_\perp\simeq\tilde\nabla$ in $f_\text{tot}$. 
Then the gradient part $F_{\rm grad}$ is given by the sum of the scalar products of 
$\tilde{\nabla}\rho^\pm$ and $\tilde{\nabla}\phi^\pm$. 
For simplicity, we neglect here the couplings between the different leaflets such as 
$(\tilde{\nabla} \rho^+)\cdot(\tilde{\nabla} \phi^-)$. 
Using again the above symmetric properties, we have
\begin{align}
F_{\rm grad} = &\frac{c}{2}\int {\rm d}^2x 
\sum_{\epsilon=+,-}\Big [(\tilde{\nabla}\rho^\epsilon)^2 
+\lambda _1(\tilde{\nabla}\phi^\epsilon)^2 \nonumber \\
& +\lambda_2(\tilde{\nabla}\rho^\epsilon)\cdot(\tilde{\nabla}\phi^\epsilon)\Big], 
\label{Fgrad}
\end{align}
where $c$ has the dimension of energy and is comparable to thermal energy, $\lambda_1$ 
and $\lambda_2$ are the dimensionless parameters of order unity.

Some comments are in order.
(i) We have thirteen parameters in our free energy; $\sigma$, $\kappa$, $k$, $d$, $\nu$, $\Lambda_i$ $(i=1,\cdots, 5)$, $c$, $\lambda_1$ and $\lambda_2$. 
In fact they all depend on the temperature $T$ and the reference densities $\psi_{\rm J0}$.
In this paper, however, we regard them as independent parameters although they cannot be varied 
independently in experiments. 
In the following sections, we investigate the behaviors of the relaxation rates as these 
parameters are varied, especially when the instability boundary of the one phase state is 
approached.

(ii) In the above total free energy $F$, terms which are purely linear in $H$ do not exist. 
They can be always eliminated by using the invariance of the system under the interchange of 
the two leaflets, which flips the sign of $H$.
Notice that the terms which are linear in $\rho_{\rm J}^\pm$ have already been taken into 
account in the definition of the surface tension $\sigma$ in eq.~(\ref{sigma}).

(iii) In principle, the free energy can include terms linear in Gaussian curvature 
$K$ which is proportional to $h^2$. 
However, without any topological change of the membrane, the integral of $K$ depends only 
on the geodesic curvature along the boundary of the membrane. 
As long as the topology and the geodesic curvature at the edge of the membrane are fixed, 
the integral merely adds a constant to the free energy. 
For this reason, we do not include any Gaussian curvature term in eq.~(\ref{Fdef}).

%%%%%%%%%%%%%%%%%%%%%%%%%%%%%%%%%%%
\subsection{Fourier representation}

The in-plane Fourier transform of any function $g(\tilde{\bm x})$ in the monolayer is defined by
\begin{equation}
g(\tilde{\bm q})=\int {\rm d}^2x \, g(\tilde{\bm x}) e ^{-i\tilde{\bm q}\cdot \tilde{\bm x}},  
\label{Fourier}
\end{equation}
where $\tilde{\bm x}= (x,y)$ and $\tilde{\bm q} =(q_x, q_y)$.
It is convenient to introduce the following new variables 
\begin{align}
&\rho= (\rho^+-\rho^-)/2, \quad \bar\rho= (\rho^++\rho^-)/2 \label{mrho}, \\
&\phi= (\phi^+-\phi^-)/2, \quad \bar\phi= (\phi^++\phi^-)/2 \label{mphi}, \\
&\hat h= h/d, 
\label{hhat}
\end{align}
and define the column vectors
\begin{equation}
{\bm a}= (\hat{h}, \rho, \phi)^{\rm T}, \quad 
{\bm b}= (\bar\rho, \bar\phi)^{\rm T}, 
\label{ab}
\end{equation}
where ``T" denotes the transpose.

The total free energy is alternatively expressed in term of the Fourier modes as
\begin{equation}
F=\int \frac{{\rm d}^2q}{(2\pi)^2} \frac{1}{2} \left[ {\bm a}^\dag A{\bm a} 
+{\bm b}^\dag B{\bm b} \right], 
\label{FFourier}
\end{equation}
where $\dag$ denotes the conjugate transpose.
In the above, $A$ and $B$ are symmetric matrices of $3\times 3$ and $2\times 2$, respectively. 
Owing to the rotational symmetry, their components depend only on the magnitude of 
the wave vector, $q = \vert \tilde{\bm q} \vert$, and are given by
\begin{align}
&A_{11}=\sigma d^2 q^2 +(\kappa+kd^2\Omega_0)d^2 q^4,  \label{A11}\\
&A_{12}=A_{21}=-kd^2 \Omega_1 q^2,  \label{A12}\\
&A_{13}=A_{31}=-kd^2  \Omega_2 q^2,   \label{A13}\\
&A_{22}=k(2-\Lambda_3)+2cq^2, \label{A22}\\
&A_{23}=A_{32}=k(\Lambda_2-\Lambda_5) +c \lambda_2  q^2,  \label{A23}\\
&A_{33}=k(2\Lambda_1-\Lambda_4) +2c\lambda_1  q^2,  \label{A33}
\end{align}
and
\begin{align}
&B_{11}=k(2+\Lambda_3)+2c q^2,  \label{B11} \\
&B_{12}=B_{21}=k(\Lambda_2+\Lambda_5) +c \lambda_2 q^2,  \label{B12}\\
&B_{22}=k(2\Lambda_1+\Lambda_4) +2c \lambda_1  q^2. \label{B22}
\end{align}
Here we have introduced the following dimensionless combinations
\begin{align}
&\Omega_0= 2+2\nu^2\Lambda_1+2\nu\Lambda_2-\Lambda_3-\nu^2\Lambda_4-2\nu\Lambda_5, \label{Omega0}\\
&\Omega_1= 2+\nu\Lambda_2-\Lambda_3-\nu\Lambda_5, \label{Omega1}\\
&\Omega_2= 2\nu\Lambda_1+\Lambda_2-\nu\Lambda_4-\Lambda_5. \label{Omega2}
\end{align}

It is important to note that ${\bm a}$ and ${\bm b}$ are decoupled in eq.~(\ref{FFourier}). 
This is due to the symmetry of the system under the interchange of the two monolayers, 
{\it i.e.}, ${\bm a}$ changes its sign under this interchange while ${\bm b}$ does not.

%%%%%%%%%%%%%%%%%%%%%%%%%%%
\section{Dynamic equations}
%%%%%%%%%%%%%%%%%%%%%%%%%%%
\label{dynamics}

In this section, we present the dynamic equations for a two-component bilayer
membrane surrounded by a viscous fluid.
We shall take into account (i) the flows in the surrounding fluid and in the membrane, 
(ii) the frictional force between the two monolayers, and (iii) the 
mutual diffusion in each monolayer. 
The surrounding fluid is assumed to be incompressible, while the membrane itself is 
compressible~\cite{Seifert}. 
Our dynamic equations are based on the standard irreversible thermodynamics~\cite{degroot,Landau}, 
and ensure that the dissipation in the whole system is non-negative definite (see Appendix \ref{appa}).
While our derivation presented in this section is self-contained, they can be formulated in 
a more systematic manner by using the so called Onsager's variational principle 
(see Appendix \ref{appb})~\cite{Onsager1, Onsager2, Doi}.

%%%%%%%%%%%%%%%%%%%%%%%%%%%%%%%%%%%
\subsection{Hydrodynamic equations}

We use ${\bm v}$ to denote the velocity field of the surrounding fluid which is assumed to be 
incompressible and to have a low Reynolds number. 
Then ${\bm v}$ for $z>0$ and $z<0$ obeys the Stokes equation
\begin{equation}
\eta \nabla^2 {\bm v}-\nabla p=0,
\label{Stokes}
\end{equation}
where $\eta$ is the shear viscosity, $\nabla=(\partial_x, \partial_y, \partial_z)$ the nabla 
operator in 3D space, and $p$ the pressure of the fluid that is determined by the 
incompressibility condition
\begin{equation}
\nabla \cdot {\bm v}=0. 
\label{incompressible}
\end{equation}

Let $\tilde {\bm v}_{\rm J}^\pm$ denote the flow velocity of the lipid ${\rm J}$ 
in the upper ($+$) and the lower ($-$) monolayers. 
Here the flow velocity is defined as the lipid mass flux divided by the mass 
density $\psi_{\rm J}^\pm$. 
We consider the dynamic equations only within the linear order in 
$\tilde{\bm v}_{\rm J}^\pm$, $h$ and $\rho_{\rm J}^\pm$. 
The average lipid velocities $\tilde{\bm v}^\pm$ in the upper and lower monolayers are 
defined as 
\begin{equation}
\tilde {\bm v}^\pm= \frac{\psi_{\rm A}^\pm \tilde {\bm v}_{\rm A}^\pm+
\psi_{\rm B}^\pm \tilde{\bm v}_{\rm B}^\pm}{\psi_{\rm A}^\pm+\psi_{\rm B}^\pm},
\end{equation}
which can be approximated within the linear order as 
\begin{equation}
\tilde{\bm v}^\pm=\frac{\psi_{\rm A0} \tilde {\bm v}_{\rm A}^\pm+\psi_{\rm B0}\tilde  {\bm v}_{\rm B}^\pm}{\psi_{\rm A0}+\psi_{\rm B0}}.
\label{vlayer}
\end{equation}

The diffusive flux of lipid A is given by
\begin{equation}
{\bm j}_{\rm d}^\pm = \psi_{\rm A0}(\tilde{\bm v}_{\rm A}^\pm-\tilde{\bm v} ^\pm) 
=-\psi_{\rm B0}(\tilde{\bm v}_{\rm B}^\pm-\tilde{\bm v} ^\pm),
\end{equation}
where use has been made of eq.~(\ref{vlayer}) in the second equality. 
It should be noted here that the diffusive flux of lipid B is given by $-{\bm j}_{\rm d}^\pm$. 
Then the continuity equations for the lipids A and B,     
$\partial\psi_{\rm J}^\pm/\partial t=-\tilde\nabla\cdot(\psi_{\rm J}^\pm \tilde{\bm v}_{\rm J}^\pm)$,
can be approximated as 
\begin{align}
&\frac{\partial \rho_{\rm A}^\pm}{\partial t} = 
-\tilde\nabla \cdot \tilde {\bm v}^\pm -\frac{1}{\psi_{\rm A0}}\tilde\nabla \cdot {\bm j}_{\rm d}^\pm,
\label{Acon}\\
&\frac{\partial \rho_{\rm B}^\pm}{\partial t} = -\tilde\nabla \cdot \tilde {\bm v}^\pm +\frac{1}{\psi_{\rm B0}}\tilde\nabla \cdot {\bm j}_{\rm d}^\pm.
\label{Bcon}
\end{align}
We further note that eqs.~(\ref{Acon}) and (\ref{Bcon}) can be expressed in simpler forms by using 
$\rho^\pm$ and $\phi^\pm$ as 
\begin{align}
&\frac{\partial \rho^\pm}{\partial t} = -\tilde\nabla \cdot \tilde {\bm v}^\pm,
\label{rhocon}\\
&\frac{\partial \phi^\pm}{\partial t} = -\tilde\nabla\cdot {\bm j}^\pm_\phi, 
\label{phicon}
\end{align}
where the diffusive flux associated with $\phi$ is now defined as 
${\bm j}^\pm_\phi = (\psi_{\rm A0}^{-1}+\psi_{\rm B0}^{-1}){\bm j}^\pm_{\rm d}$.

As in the standard irreversible thermodynamics, ${\bm j}^\pm_\phi$ is assumed to be 
proportional to the gradient of the effective chemical potential 
$\mu^\pm =(\mu_{\rm A}^\pm/m_{\rm A})-(\mu_{\rm B}^\pm/m_{\rm B})$, where $m_{\rm J}$ and 
$\mu^\pm_{\rm J}$ are the molecular mass and the chemical potential per molecule 
for lipid ${\rm J}$, respectively~\cite{Landau}. 
The chemical potentials are given by $\mu^\pm_{\rm J}=m_{\rm J}(\delta F/\delta \psi^\pm_{\rm J})$. 
Then the diffusive flux in eq.~(\ref{phicon}) becomes
\begin{align}
{\bm j}_\phi^\pm=- L_\phi\Big( \frac{1}{\psi_{\rm A0}}+\frac{1}{\psi_{\rm B0}}\Big)^{-1}\tilde{\nabla}\mu^\pm
=- L_\phi \tilde{\nabla} \frac{\delta F}{\delta \phi^\pm},
\label{Dflux}
\end{align}
where $L_\phi>0$ is the Onsager coefficient~\cite{degroot}, and the second equality follows 
from the relation, $\mu^\pm=\delta F/\delta\psi^\pm_{\rm A}-\delta F/\delta\psi^\pm_{\rm B}=
(\psi_{\rm A0}^{-1}+\psi_{\rm B0}^{-1}) \delta F /\delta \phi^\pm$. 
In the definition of $L_\phi$, we have intentionally put the factor 
$[(1/\psi_{\rm A0})+(1/\psi_{\rm B0})]^{-1}$ in order to make eq.~(\ref{phicon}) simpler. Equation (\ref{Dflux}) indicates that, as in usual 3D multi-component fluids, mutual diffusion occurs essentially due to the inhomogeneity of the density difference $\phi^\pm$ between the lipid A and B in each monolayer. Furthermore, even if $\phi^\pm$ are homogeneous, mutual diffusion can still be induced by the inhomogeneity of $h$ and $\rho^\pm$ that are coupled to $\phi^\pm$ via the free energy.

Next we discuss the force balance conditions. 
We regard each monolayer as a compressible 2D fluid characterized by the shear viscosity 
$\mu$ and the bulk viscosity $\zeta$. 
The 2D viscous stress tensors $\tau_{ij}^\pm$ in the monolayers are given by
\begin{equation}
\tau_{ij}^\pm=\mu (\partial_i \tilde v_j^\pm+\partial_j \tilde v_i^\pm) +
(\zeta-\mu) \delta_{ij} \tilde\nabla\cdot \tilde{\bm v}^\pm.
\label{stresslayer}
\end{equation}
On the other hand, the reversible force density due to the in-plane pressure is given by
\begin{equation}
{\bm f}^\pm=-\sum_{{\rm J}={\rm  A, B}} \psi_{\rm J0}\tilde\nabla 
\frac{\delta F}{\delta \psi_{\rm J}^\pm}=-\tilde\nabla\frac{\delta F}{\delta \rho^\pm}, 
\label{flayer}
\end{equation}
up to the linear order~\cite{Bitbol2}.
The force balance equations in the tangential direction of the monolayers are given by
\begin{equation}
f^\pm_i + \partial _j \tau_{ij}^\pm
\pm T^\pm_{iz} \mp b(\tilde v^+_i - 
\tilde v^-_i)=0, 
\label{lateralforce}
\end{equation}
for $i=x, y$. 
Here $T^\pm_{ij}$ are the stress tensors of the surrounding fluid 
$T_{ij} = -p \delta_{ij}+\eta (\partial_i v_j +\partial _j v_i )$ evaluated at 
$z\rightarrow \pm 0$. 
The last term in eq.~(\ref{lateralforce}) represents the frictional forces between the upper 
and the lower monolayers, and $b$ is the friction coefficient~\cite{Seifert,Yeung}.

In the normal $z$-direction, the restoring force of the membrane is balanced with the 
normal force due to the surrounding fluid.
Hence we have 
\begin{align}
T_{zz}^+-T_{zz}^- =\frac{\delta F}{\delta h}=
&-\sigma\tilde\nabla^2 h +(\kappa+kd^2\Omega_0)\tilde\nabla^2 \tilde\nabla^2 h  \nonumber \\
&+kd(\Omega_1 \tilde\nabla^2\rho+\Omega_2 \tilde\nabla^2\phi), 
\label{normalforce}
\end{align}
where the last expression follows from eqs.~(\ref{Fdef}), (\ref{Fcoup}) and 
(\ref{Omega0})--(\ref{Omega2}).

We further assume that the non-slip boundary condition holds at the upper and the lower monolayers. 
Let ${\bm v}^\pm=(v^\pm_x, v^\pm_y, v^\pm_z)$ denote the velocity of the surrounding 
fluid ${\bm v} $ evaluated at $z\to \pm0$.
The tangential components of ${\bm v}^\pm$ should coincide with the average velocities of 
the monolayers 
\begin{equation}
v^\pm_i=\tilde v^\pm_i, 
\label{BC1}
\end{equation}
for $i=x,y$. 
On the other hand, the normal components $v^\pm_z$ should coincide with the time derivative 
of the membrane height $h$
\begin{equation}
v^\pm_z=\frac{\partial h}{\partial t}. 
\label{BC2}
\end{equation}

Up to now, we have presented a set of dynamic equations given by 
eqs.~(\ref{Stokes}), (\ref{incompressible}), (\ref{rhocon}), (\ref{phicon}), 
(\ref{Dflux}), (\ref{lateralforce}), (\ref{normalforce}), (\ref{BC1}) and (\ref{BC2})
to be solved. 
As mentioned before, they can be also derived systematically by using the Onsager's 
variational principle explained in Appendix \ref{appb}.

%%%%%%%%%%%%%%%%%%%%%%%%%%%%%%%%%%%%%%%%%%%%%%%%%%%%%%%%
\subsection{Relaxation equations for membrane variables}
\label{relaxation}

From the derived dynamic equations, we can integrate out the velocity fields ${\bm v}$ 
and $\tilde{\bm v}^\pm$ to obtain the relaxation equations for the spatially Fourier 
transformed dynamical variables, $\rho^\pm (\tilde{\bm q},t)$, $\phi^\pm (\tilde{\bm q},t)$ 
and $\hat{h}(\tilde{\bm q},t)=h(\tilde{{\bm q}},t)/d$ (see eq.~(\ref{Fourier})).
The details are described in Appendix \ref{appc} and the resulting equations are
\begin{align}
\frac{\partial {\bm a}}{\partial t}& =-\Gamma_a (q) {\bm a} \label{Dmatrixa}, \\
\frac{\partial {\bm b}}{\partial t}& =-\Gamma_b (q) {\bm b} \label{Dmatrixb},
\end{align}
where ${\bm a}$ and ${\bm b}$ are defined in eq.~(\ref{ab}).
In the above, the matrices $\Gamma_a$ and $\Gamma_b$ are given by
\renewcommand{\arraystretch}{1.8}
\begin{equation}
\Gamma_a(q)=
\begin{pmatrix}
\displaystyle \frac{A_{11}}{4\eta d^2 q} && \displaystyle\frac{A_{12}}{4\eta d^2 q} &&\displaystyle \frac{A_{13}}{4\eta d^2 q} \\
 c_1A_{12}q^2 && c_1A_{22}q^2 && c_1A_{23}q^2  \\
\frac{1}{2} L_\phi A_{13}q^2 && \frac{1}{2} L_\phi A_{23}q^2 && \frac{1}{2}L_\phi A_{33}q^2 
\end{pmatrix},
\label{gamma_a}
\end{equation}
\renewcommand{\arraystretch}{1}
and
\renewcommand{\arraystretch}{1.8}
\begin{equation}
\Gamma_b(q)=
\begin{pmatrix}
c_2B_{11}q^2  && c_2B_{12}q^2 \\
\frac{1}{2} L_\phi B_{12}q^2  && \frac{1}{2} L_\phi B_{22}q^2  
\end{pmatrix},
\label{gamma_b}
\end{equation}
\renewcommand{\arraystretch}{1}
with
\begin{align}
&c_1 = [4b+4\eta q+2(\mu+\zeta)q^2]^{-1}, \label{c1}\\
&c_2 = [4\eta q+2(\mu+\zeta)q^2]^{-1}. \label{c2}
\end{align}
The eigenvalues of $\Gamma_a$ and $\Gamma_b$ correspond to the relaxation rates of the 
binary bilayer membranes. The equations for the five dynamical variables are split into the decoupled two equations (\ref{Dmatrixa}) and (\ref{Dmatrixb}), where eq.~(\ref{Dmatrixa}) changes its sign under the interchange of the two monolayers, while eq.~(\ref{Dmatrixb}) does not. This is the consequence of the symmetry of the hydrodynamic equations as well as that of the free energy (the latter is discussed after eq.~(\ref{Omega2})).
In the next section, we will examine how these relaxation rates behave as the coupling 
parameters $\Lambda_i$ are varied.

%%%%%%%%%%%%%%%%%
\section{Results}
%%%%%%%%%%%%%%%%%
\label{results}

%%%%%%%%%%%%%%%%%%%%%%%%%%%%%
\subsection{Parameter values}

In Table~\ref{TabPara}, we list the set of parameter values chosen in our numerical calculations.
Following previous experiments \cite{Helfrich, Song, Rawicz}, the bending modulus $\kappa$ is set equal to $10^{-12}$ erg.  As discussed after eq.~(\ref{alphabeta}), the lengths $d$ and $d\nu$ are comparable with the monolayer thickness. Then we may set $d=10^{-7}$ cm, with $\nu$ of order unity. The combination $c\lambda_i$ ($i=1,2$) in $F_{\rm grad}$ is related to the line tension $\xi$ in a phase separated membrane as $\xi \sim (k_{\rm B}Tc)^{1/2}\lambda_i/d$. Since $\xi$ has been measured to be several pN \cite{Tian}, we may set $c=10^{-14}$ ${\rm erg}$, with $\lambda_i$ of order unity. 
The surface tension $\sigma$ can take extremely wide range of values depending on 
experimental conditions.
For vesicles in a solution, it can be controlled by changing the osmotic pressure 
difference between the inside and outside of the vesicles.
In the following, we will examine two cases, namely, 
the small tension case with $\sigma=10^{-8}$ ${\rm erg / cm}^{2}$ and 
the moderate tension case with $\sigma=10^{-4}$ ${\rm erg / cm}^{2}$. 
The coefficients $A_{22}$ and $B_{11}$ in eqs.~(\ref{A22}) and (\ref{B11}) can be interpreted as the moduli associated with the total densities in the upper and the lower monolayers, respectively (to be more precise, the moduli of their linear combinations $\rho$ and $\bar\rho$ in eq.~(\ref{mrho})). Then $k(2-\Lambda_3)$ and $k(2+\Lambda_3)$ are comparable with the areal compression moduli. Following previous experiments \cite{Evans1, Evans2}, we set $k=70$ erg/cm$^2$ with $|\Lambda_3|\ll1$.

The remaining parameters that have yet to be determined in the free energy are $\Lambda_1$, $\Lambda_2$, $\Lambda_4$ and $\Lambda_5$. These parameters depend on the temperature and the average composition. We find in the following, however, that the behavior of the decay rates is not sensitive to these parameters, unless the reduced temperatures $\tau_a$ and $\tau_b$ defined below in eqs.~(\ref{taua}) and (\ref{taub}) are very close to zero. When they are close to zero (but positive), the associated diffusive modes become extremely slow. This point will be discussed later in more detail.

We next discuss the kinetic parameters.
The membrane viscosities $\mu$ and $\zeta$ appear only as a sum $\mu+\zeta$ in 
$\Gamma_a$ and $\Gamma_b$.
Since we could not find any reliable value of $\zeta$ in the literatures, we set 
$\mu+\zeta=10^{-7}$ ${\rm erg\cdot s/cm^2}$.
(The membrane bulk viscosity $\zeta$ was neglected in ref.~\cite{Seifert}.)
The Onsager coefficient $L_\phi$ for the mutual diffusion is roughly estimated as follows. 
Assuming that the mutual diffusion constant is on the same order as the self-diffusion 
constant $D$ of a lipid molecule, we have $D\sim (\Gamma_a)_{33}/q^2\sim L_\phi k$ 
(see eqs.~(\ref{A33}), (\ref{Dmatrixb}) and (\ref{gamma_a})).
Using the value $D\sim 10^{-7}$ ${\rm cm^2/s}$~\cite{Membook}, we obtain 
$L_\phi = 1.4\times 10^{-9}$ ${\rm cm^4 / (erg \cdot s)}$.
Several authors have reported different values of the friction coefficient $b$ \cite{Pott,Shkulipa,JBprl,Bitbol1,Merkel,Horner}. Since they are in the range of $10^{7}$ -- $3\times 10^8$ $\mathrm{erg\cdot s/cm}^4$, we set $b=2\times 10^7$ $\mathrm{erg\cdot s/cm}^4$ in this paper.

\begin{table}[tbh]
\caption{
List of static and dynamic parameters taken from the 
literatures~\cite{Seifert,Rawicz,Pott,Tian, Evans1, Evans2,Shkulipa,JBprl,Bitbol1,Helfrich, Membook, Song, Merkel,Horner} and 
used in sect.~\ref{results}.
\label{TabPara}}
\begin{ruledtabular}
\begin{tabular}[t]{c c c c c}
$\sigma$  \quad & $\kappa$\ \ \quad  & $k$\   \quad & $d$\  \quad & $c$\\
 $\mathrm{erg/cm}^2\quad$ & $\mathrm{erg}$\quad & $\mathrm{erg}/\mathrm{cm}^2$ \quad & $\mathrm{cm}$\quad & $\mathrm{erg}$\\
\hline
$10^{-8}$ or $10^{-4}$ & $10^{-12} $ & $70$ & $10^{-7} $ \ & $10^{-14}$\\
\end{tabular}
\end{ruledtabular}
\vspace{0mm}
\begin{ruledtabular}
\begin{tabular}[t]{c c c c}
$\eta$ & $b$\  \quad  & $\mu +\zeta $ \quad & $L_\phi$ \quad  \\
$\mathrm{erg}\cdot \mathrm{s}/\mathrm{cm}^3\quad $&$\mathrm{erg}\cdot \mathrm{s}/\mathrm{cm}^4\quad$&$\mathrm{erg}\cdot \mathrm{s}/\mathrm{cm}^2$ &$\mathrm{cm}^4 / (\mathrm{erg} \cdot\mathrm{s})$\\
\hline
$10^{-2}$&$2\times 10^{7}$&$10^{-7}$& $1.4\times 10^{-9}$ \\
\end{tabular}
\end{ruledtabular}
\end{table}

%%%%%%%%%%%%%%%%%%%%%%%%%%%%%%%%%
\subsection{Stability conditions}

The wave number dependent susceptibilities $\chi_a(q)$ and $\chi_b(q)$ are defined as 
the reciprocals of the eigenvalues of the matrices $A$ and $B$ in eq.~(\ref{FFourier}),
respectively.
Then the thermodynamic stability of the one phase state (without any phase separation)
is ensured when these susceptibilities are all positive. 
Since $A$ and $B$ are $3\times 3$ and $2\times 2$ matrices, respectively, there are 
three $\chi_a$ and two $\chi_b$ values which can be explicitly obtained in principle.
However, since their full expressions are tedious, we discuss here the 
conditions for the thermodynamic stability at $q=0$ and $\infty$. 
More detailed discussions are given in Appendix \ref{appd} where we also show that the 
instability characterized by intermediate wave numbers does not occur as 
long as the stability conditions at $q=0$ and $\infty$ are satisfied.

\begin{figure}[tbh]
\includegraphics[scale=0.39]{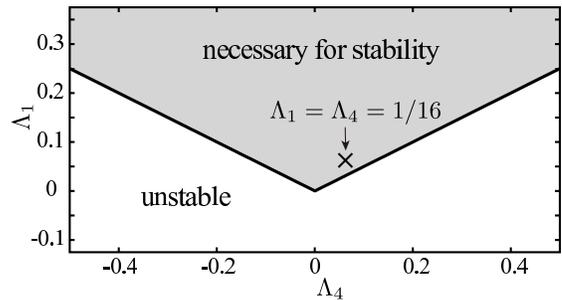}
\caption{
Stability diagram of the one phase state at $q=0$ in the $(\Lambda_1, \Lambda_4)$-plane
as expressed by eq.~(\ref{stab2}).
For the thermodynamic stability of the one phase state, $\Lambda_1$ and $\Lambda_4$ need to be 
in the gray region.
The cross corresponds to the parameter values $\Lambda_1= \Lambda_4=1/16$ which we use in 
fig.~\ref{FigDiagram2} and in the numerical analysis in figs.~\ref{FigrateAlow1}--\ref{FigrateB2}.} 
\label{FigDiagram1}
\end{figure}

As $q\to \infty$, we find that the susceptibilities $\chi_a$ and $\chi_b$ are both positive 
if and only if 
\begin{equation}
0<\lambda_1-\frac{\lambda_2^2}{4}\equiv \Delta_\lambda. 
\label{stab_highQ}
\end{equation}
Hereafter we assume that the above condition is always satisfied. 
The stability at $q=0$, on the other hand, is ensured by the positivity of $\chi_a(0)$ 
and $\chi_b(0)$, which is realized if and only if 
\begin{align}
&|\Lambda_3|<2, 
\label{stab1}\\
&0<\Lambda_1, \quad |\Lambda_4|<2\Lambda_1, 
\label{stab2}
\end{align}
and
\begin{align}
&|\Lambda_2-\Lambda_5|<[(2-\Lambda_3)(2\Lambda_1-\Lambda_4)]^{1/2}, \label{stab3}\\
&|\Lambda_2+\Lambda_5|<[(2+\Lambda_3)(2\Lambda_1+\Lambda_4)]^{1/2}. \label{stab4}
\end{align}
The conditions eqs.~(\ref{stab1}) and (\ref{stab2}) are equivalent to 
$A_{22}$, $A_{33}$, $B_{11}$, $B_{22}>0$ at $q=0$. 
In fig.~\ref{FigDiagram1}, we plot the condition eq.~(\ref{stab2}) on the 
$(\Lambda_1, \Lambda_4)$-plane.  
For the stability of the one phase state, $\Lambda_1$ and $\Lambda_4$ need to 
be within the gray region. 
Otherwise the system is unstable towards the phase separation.
Given that $\Lambda_1$ and $\Lambda_4$ are fixed at the values satisfying eq.~(\ref{stab2}), 
the stability conditions for $\Lambda_2$, $\Lambda_3$ and $\Lambda_5$ are given by 
eqs.~(\ref{stab1}), (\ref{stab3}) and (\ref{stab4}).

\begin{figure}[tbh]
\includegraphics[scale=0.48]{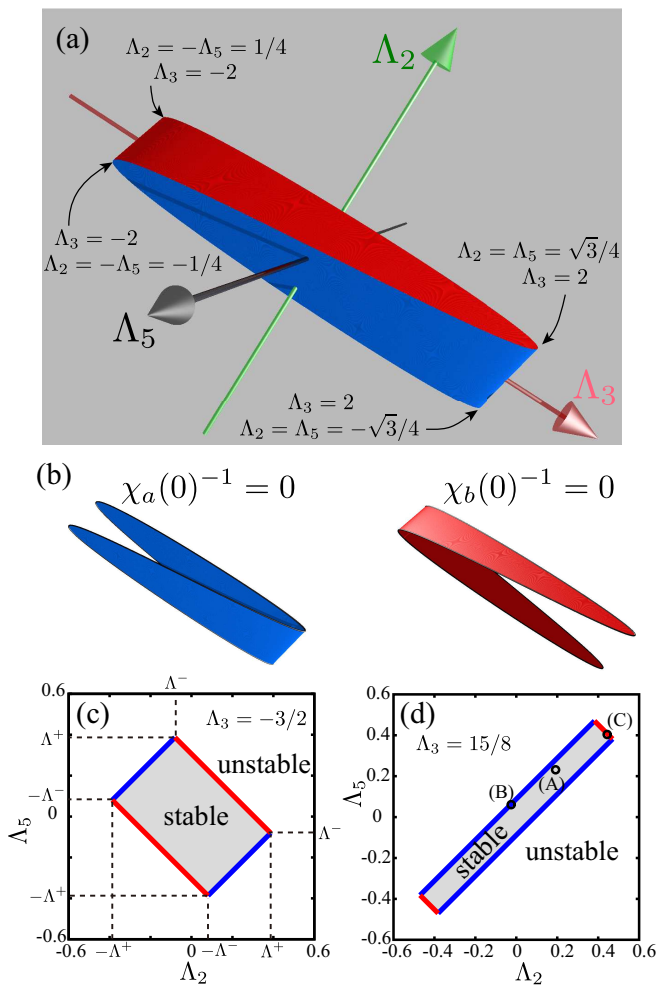}
\caption{(a) Stability diagram of the one phase state at $q=0$ in the 
$(\Lambda_2, \Lambda_3, \Lambda_5)$-space when $\Lambda_1= \Lambda_4=1/16$ 
as marked in fig.~\ref{FigDiagram1}.
The one phase state is stable if $\Lambda_2$, $\Lambda_3$ and $\Lambda_5$ are inside the 
region enclosed by the surface.
(b) The enclosing surface consists of two surfaces; the blue one at which $1/\chi_a(0)=0$ 
and the red one at which $1/\chi_b(0)=0$. 
The cross sections of the stable region on the $(\Lambda_2,\Lambda_5)$-plane 
when (c) $\Lambda_3=-3/2$ and (d) $\Lambda_3=15/8$.
The circles marked with (A)--(C) in (d) correspond to the parameter values used in 
figs.~\ref{FigrateAlow1}--\ref{FigTimeMode2}.
} 
\label{FigDiagram2}
\end{figure}

\begin{figure}[tbh]
\includegraphics[scale=0.32]{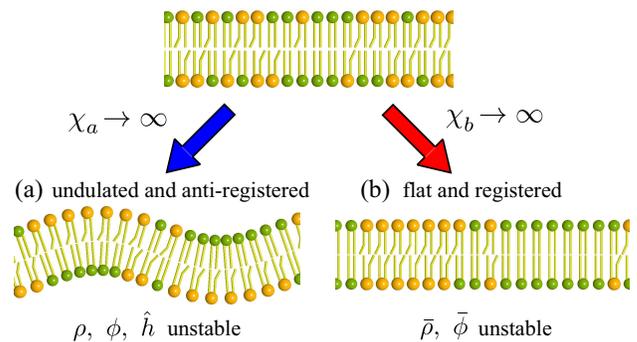}
\caption{Schematic illustrations of the two possible instabilities. 
Note that for the sake of clarity (i) the difference in the lipid heights is not 
drawn, and (ii) a strong phase separation is represented. 
In this paper we investigate, however, only the homogeneous phase in the vicinity 
of the phase separation. 
(a) Anti-registered instability. 
As one of the $\chi_a$ values diverges, the density difference 
between the two monolayers and the bending mode become unstable.
(b) Registered instability. 
As one of the $\chi_b$ values diverges, the sum of the densities 
in the upper and lower monolayers becomes unstable.}
\label{FigSchematic}
\end{figure}

In fig.~\ref{FigDiagram2}(a), we show the stable region in the 
$(\Lambda_2, \Lambda_3,\Lambda_5)$-space when $\Lambda_1=\Lambda_4=1/16$ as marked 
by a cross in fig.~\ref{FigDiagram1}. 
The stable region is enclosed by a surface which consists of blue and red parts.
On the blue surface (fig.~\ref{FigDiagram2}(b) left), one of the three 
$\chi_a$-values diverges at $q=0$ and its corresponding mode becomes unstable, whereas on the 
red surface (fig.~\ref{FigDiagram2}(b) right), one of the two $\chi_b$-values 
diverges at $q=0$.
One can confirm from eqs.~(\ref{stab3}) and (\ref{stab4}) that the cross section 
of the stable region on the $(\Lambda_2, \Lambda_5)$-plane at constant $\Lambda_3$  
is given by an oblique rectangle whose center is at $\Lambda_2=\Lambda_5=0$. 
In Figs.~\ref{FigDiagram2}(c) and (d), we present the cross sections at $\Lambda_3=-3/2$ and 
$15/8$, respectively. 
As shown in (c), the apex coordinates of the rectangle $\Lambda^\pm$ are given by
\begin{align}
\Lambda^\pm  = &\frac{1}{2} \left[ \sqrt{(2+\Lambda_3)(2\Lambda_1+\Lambda_4)} \right.
\nonumber \\
& \left. \pm \sqrt{(2-\Lambda_3)(2\Lambda_1-\Lambda_4)} \right].
\end{align}
These values are $\Lambda^\pm \approx 0.153\pm0.234$ and 
$\Lambda^\pm \approx 0.426\pm0.0442$ in fig.~\ref{FigDiagram2}(c) and (d), respectively.

As $\Lambda_3$ is increased towards $2$, the blue sides of the cross section become 
longer while the red sides become shorter.
In the limit of $\Lambda_3\nearrow 2$, the stable region eventually turns out to be 
a line segment whose endpoints are given by $(\Lambda_2, \Lambda_5)=(\pm\sqrt{2\Lambda_1+\Lambda_4}, 
\pm\sqrt{2\Lambda_1+\Lambda_4})=(\pm \sqrt{3}/4, \pm\sqrt{3}/4)$. 
In the limit of $\Lambda_3\searrow -2$, on the other hand, the stable region shrinks to 
a line segment with the endpoints at $(\Lambda_2, \Lambda_5)=(\pm\sqrt{2\Lambda_1-\Lambda_4}, 
\mp\sqrt{2\Lambda_1-\Lambda_4})=(\pm 1/4, \mp1/4)$.

Even if we choose other $\Lambda_1$ and $\Lambda_4$ values in the stable 
region of fig.~\ref{FigDiagram1}, the qualitative features of the stable region in the 
$(\Lambda_2, \Lambda_3, \Lambda_5)$-space remains the same. 
However, as the combination $(\Lambda_1, \Lambda_4)$ approaches the boundary of the gray region, 
the stable region in the $(\Lambda_2, \Lambda_3, \Lambda_5)$-space becomes narrower, and 
eventually disappears just at the boundary of the stable region.
In fig.~\ref{FigSchematic}, we illustrate the corresponding instabilities to take place.
When one of the $\chi_a$-values diverges, a certain linear combination of $\rho$, $\phi$ and 
$\hat{h}$ becomes unstable as in fig.~\ref{FigSchematic}(a), while a linear combination of 
$\bar\rho$ and $\bar\phi$ becomes unstable as in  fig.~\ref{FigSchematic}(b) when one of 
the $\chi_b$-values diverges. 
Hereafter we shall call the instabilities of type (a) and (b) 
the ``anti-registered instability" and the ``registered instability", respectively. Note that these two types of instabilities are purely the consequences of the symmetry of the system (see also the sentences after eq.~(\ref{Omega2})).

So far, we have discussed the stability conditions of the one phase state at $q=0$ and 
$\infty$.
In Appendix \ref{appd}, we show that the instability does not take place for intermediate 
wave numbers as long as the membrane is stable at $q=0$ and $\infty$.
Hence the stability conditions are generally given by eqs.~(\ref{stab1})--(\ref{stab4}).

%%%%%%%%%%%%%%%%%%%%%%%%%%%%%
\subsection{Relaxation rates}

\begin{figure}[tbh]
\includegraphics[scale=0.7]{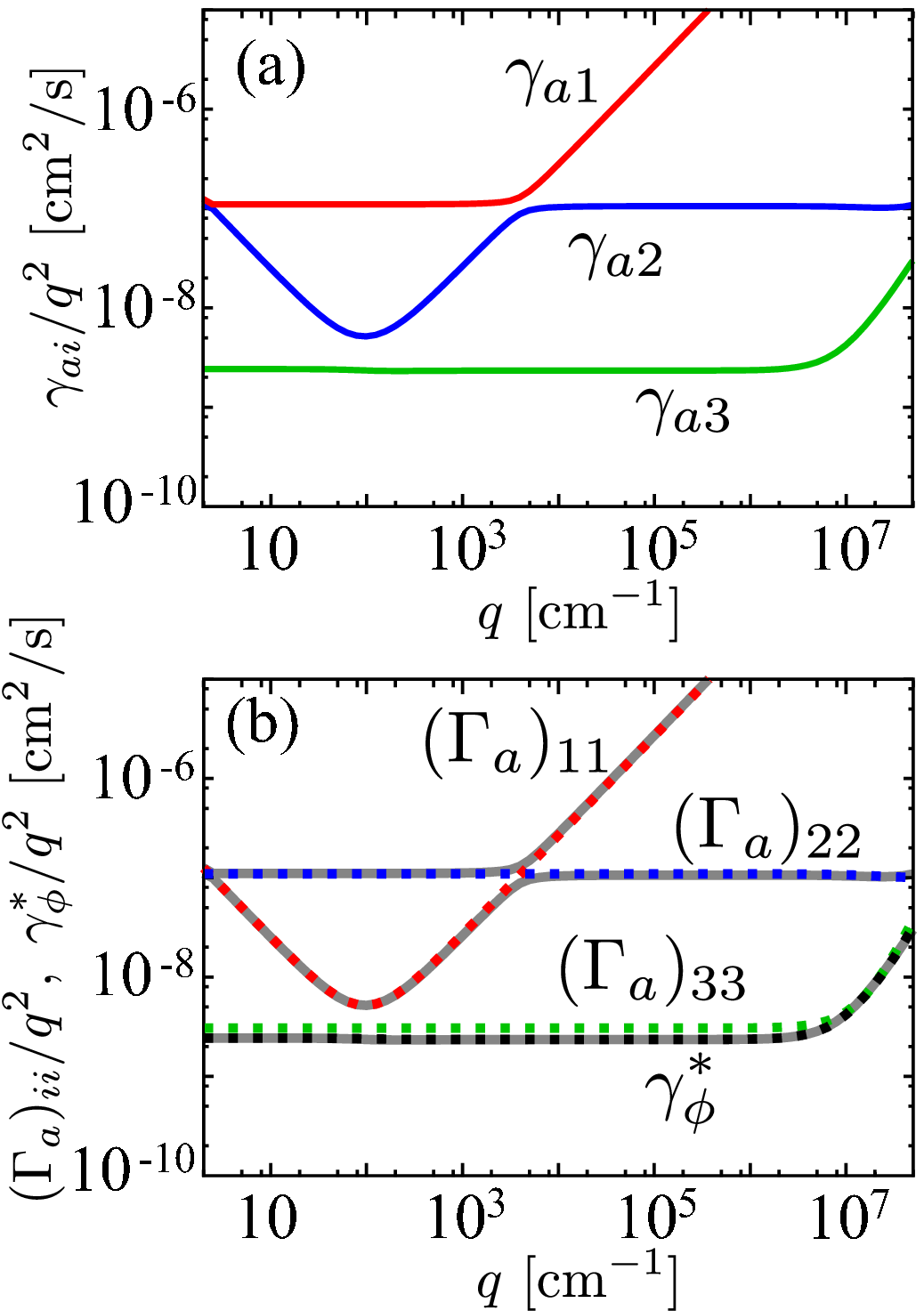}
\caption{Eigenmodes of $\Gamma_a$ for small tension case; 
$\sigma=10^{-8}$ $\mathrm{erg}/\mathrm{cm}^2$.
The parameter values are $(\Lambda_2, \Lambda_5)=(0.193, 0.233)$ as 
marked (A) in fig.~\ref{FigDiagram2}(d), and the membrane is not close to the 
anti-registered instability boundary.
(a) Plots of the relaxation rates $\gamma_{ai}$ ($i=1,2,3$) as a function of the 
wave number $q$.
(b) Plots of the diagonal elements $(\Gamma_a)_{ii}$ ($i=1,2,3$) of the matrix $\Gamma_a$ 
as a function of the wave number $q$ (dashed color lines).
The effective decay rate $\gamma_\phi^*$ is plotted with a black dashed line. 
For comparison, the relaxation rates $\gamma_{ai}$ in (a) are also plotted with 
grey solid lines.
For convenience, all the plotted quantities are divided by $q^2$.
} 
\label{FigrateAlow1}
\end{figure}

\begin{figure}[tbh]
\includegraphics[scale=0.7]{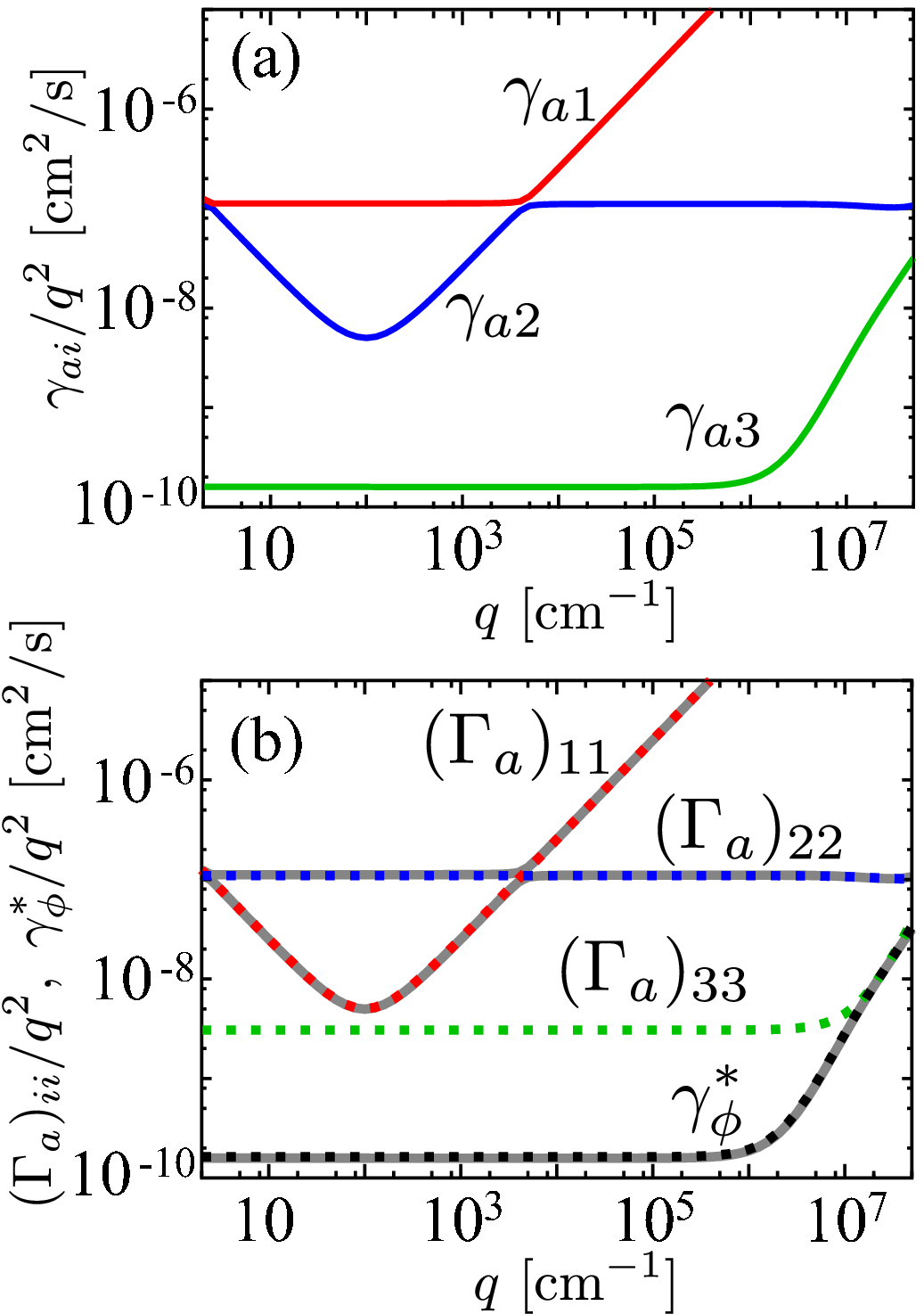}
\caption{Eigenmodes of $\Gamma_a$ for small tension case; 
$\sigma=10^{-8}$ $\mathrm{erg}/\mathrm{cm}^2$.
The parameter values are $(\Lambda_2, \Lambda_5)=(-0.023, 0.063)$ as 
marked (B) in fig.~\ref{FigDiagram2}(d), and the membrane is close to the 
anti-registered instability.
(a) Plots of the relaxation rates $\gamma_{ai}$ as a function of the 
wave number $q$.
(b) Plots of the diagonal elements $(\Gamma_a)_{ii}$ of the matrix $\Gamma_a$ 
as a function of the wave number $q$ (dashed color lines).
The effective decay rate $\gamma_\phi^*$ is plotted with a black dashed line. 
For comparison, the relaxation rates $\gamma_{ai}$ in (a) are also plotted with 
grey solid lines.
For convenience, all the plotted quantities are divided by $q^2$.
}
\label{FigrateAlow2}
\end{figure}

As a main result of this paper, we next examine the relaxation rates (or decay rates) 
of the various hydrodynamic modes in the one phase state.
They are obtained from the eigenvalues of $\Gamma_a$ and $\Gamma_b$
in eqs.~(\ref{Dmatrixa}) and (\ref{Dmatrixb}), respectively.
In the following calculations, we set the parameter values to 
$\Lambda_1=\Lambda_4=1/16$ and $\Lambda_3=15/8$ as in fig.~\ref{FigDiagram2}(d), 
while $(\Lambda_2, \Lambda_5)$ are varied. 
For simplicity, we further set $\lambda_1=\lambda_2=\nu=1$ which also satisfy 
eq.~(\ref{stab_highQ}).

%%%%%%%%%%%%%%%%%%%%%%%%%%%%%%%%%%%%%%%%%%%%%%%%%%%%%%%%%%%%
\subsubsection{Eigenmodes of $\Gamma_a$: small tension case}

Setting the parameters as $(\Lambda_2, \Lambda_5)=(0.193, 0.233)$, we plot 
in fig.~\ref{FigrateAlow1}(a) the three eigenvalues of $\Gamma_a$
denoted by $\gamma_{ai}$ ($i=1,2,3$ and $\gamma_{a1}>\gamma_{a2}>\gamma_{a3}$), 
and in (b) the three diagonal elements
of $\Gamma_a$ denoted by $(\Gamma_a)_{ii}$ ($i=1,2,3$ and 
$(\Gamma_a)_{11}> (\Gamma_a)_{22} >(\Gamma_a)_{33}$) for a small 
surface tension, $\sigma=10^{-8}$ $\mathrm{erg}/\mathrm{cm}^2$.
Similar plots are given in fig.~\ref{FigrateAlow2} when $(\Lambda_2, 
\Lambda_5)=(-0.023, 0.063)$ with the same surface tension value.  
These choices of the parameters are marked with (A) and (B) in fig.~\ref{FigDiagram2}(d). 
The system is far from and close to the unstable region in 
figs.~\ref{FigrateAlow1} and \ref{FigrateAlow2}, respectively. 
In the latter case, at least one of the eigenvalues of $A$ becomes very 
small, and the anti-registered instability shown in fig.~\ref{FigSchematic}(a) 
is about to take place.

In both figs.~\ref{FigrateAlow1} and \ref{FigrateAlow2}, the fastest decay rate 
$\gamma_{a1}$ is found to be
\begin{equation}
\gamma_{a1}\simeq 
\left\{
\begin{array}{l}
(\Gamma_a)_{22}\quad (q\ll q_{\rm mc}), \\
(\Gamma_a)_{11}\quad (q\gg q_{\rm mc}).
\end{array}
\right. 
\label{gammaA1low}
\end{equation}
Here the mode crossing wave number is given by 
\begin{equation}
q_{\rm mc}=\frac{\eta k(2-\Lambda_3)}{\kappa b}, \label{qmc}
\end{equation}
at which $(\Gamma_a)_{11}= (\Gamma_a)_{22}$ holds. 
The positivity of $q_{\rm mc}>0$ follows from the stability condition eq.~(\ref{stab1}).
Using the present parameter values, 
we obtain $q_{\rm mc} =4.38\times10^3$ $\mathrm{cm}^{-1}$.

Let us introduce the ``quasi-equilibrium" state of $\rho$ for given $\hat{h}$ and $\phi$ as 
\begin{equation}
\rho_{\rm e}^{(2)}(\hat{h},\phi, q)=-\frac{A_{12}\hat{h}+A_{23}\phi}{A_{22}}. 
\label{rhoe2}
\end{equation}
We use the term ``quasi-equilibrium" because $\rho_{\rm e}^{(2)}(\hat{h},\phi, q)$ minimizes 
the free energy $F$ under the condition that the other variables are fixed at $(\hat{h}, \phi)$. 
It can be obtained by equating the second row of eq.~(\ref{Dmatrixa}) to zero, 
and solving for $\rho$.
Then we can rewrite the second row of eq.~(\ref{Dmatrixa}) as 
$\partial \rho /\partial t=-(\Gamma_a)_{22} (\rho-\rho_{\rm e}^{(2)})$. 
Hence $\gamma_{a1} \simeq (\Gamma_a)_{22}$ for $q\ll q_{\rm mc}$ indicates that $\rho$ relaxes 
towards the quasi-equilibrium state $\rho_{\rm e}^{(2)}$ with the decay rate $\gamma_{a1}$,
while the other variables $\hat{h}$ and $\phi$ are almost unchanged (frozen) during this process.
In other words, $\rho$ relaxes much faster than $\hat{h}$ and $\phi$. 
In this regime, we can approximate $A_{22}\simeq k(2-\Lambda_3)$ and $c_1\simeq 1/(4b)$ 
in eqs.~(\ref{A22}) and (\ref{c1}), respectively, and the decay rate scales 
as $\gamma_{a1}\simeq (\Gamma_a)_{22} \simeq k(2-\Lambda_3)q^2/(4b) \sim q^2$.

Similarly, the decay rate $\gamma_{a1}$ for $q\gg q_{\rm mc}$ corresponds to the 
relaxation of $\hat{h}$ to its quasi-equilibrium state
\begin{equation}
\hat{h}_{\rm e}^{(2)}(\rho,\phi,q)=-\frac{A_{12}\rho+A_{13}\phi}{A_{11}}, 
\label{he2}
\end{equation}
while both $\rho$ and $\phi$ are frozen during the relaxation of $\hat{h}$. 
For $q\gg q^*$ with 
\begin{equation}
q^*=\sqrt{\frac{\sigma}{\kappa}}, \label{qstar}
\end{equation}
one can approximate eq.~(\ref{A11}) as $A_{11}\simeq(\kappa+kd^2\Omega_0)d^2q^4$. 
For the parameter values used in figs.~\ref{FigrateAlow1} and \ref{FigrateAlow2}, we have 
$q_{\rm mc}\gg q^*$ because $q^*=100$ ${\rm cm}^{-1}$ (see also the sentences below 
eq.~(\ref{sigmac})).
Then the decay rate scales as $\gamma_{a1}\simeq (\Gamma_a)_{11}\simeq 
(\kappa+kd^2\Omega_0)q^3/(4\eta) \sim q^3$ for $q\gg q_{\rm mc}$.

The second fastest decay rate $\gamma_{a2}$ behaves as
\begin{equation}
\gamma_{a2}\simeq
\left\{
         \begin{array}{l }
         \displaystyle (\Gamma_a)_{11}-\frac{(\Gamma_a)_{12}(\Gamma_a)_{21}}{(\Gamma_a)_{22}} \simeq (\Gamma_a)_{11}\quad (q\ll q_{\rm mc}), \\
           \displaystyle (\Gamma_a)_{22}-\frac{(\Gamma_a)_{12}(\Gamma_a)_{21}}{(\Gamma_a)_{11}}\simeq (\Gamma_a)_{22}\quad (q\gg q_{\rm mc}).
         \end{array}
    \right. \label{gammaA2low}
\end{equation}
Let us introduce the quasi-equilibrium states of $\hat{h}$ and $\rho$ for given $\phi$ as
\begin{align}
&\hat{h}_{\rm e}^{(1)}(\phi, q)=\frac{ A_{12} A_{23}- A_{13} A_{22}}{ A_{11}A_{22}- A_{12}^2} \phi \label{he1}, \\
&\rho_{\rm e}^{(1)}(\phi, q)=\frac{ A_{12} A_{13}- A_{11} A_{23}}{ A_{11}A_{22}- A_{12}^2} \phi,
\label{rhoe1}
\end{align}
which minimize the total free energy $F$ under the condition that $\phi$ is fixed.
They are obtained by equating the first and the second rows of eq.~(\ref{Dmatrixa}) to zero, 
and solving simultaneously for $\hat{h}$ and $\rho$. 
Assuming that the relaxation of $\rho$ is much faster than that of $\hat{h}$, we substitute 
$\rho\simeq\rho_{\rm e}^{(2)}$ given by eq.~(\ref{rhoe2}) into the first row of 
eq.~(\ref{Dmatrixa}) to obtain 
\begin{align}
\frac{\partial \hat{h}}{\partial t}&\simeq- \Big[  (\Gamma_a)_{11}-\frac{(\Gamma_a)_{12}(\Gamma_a)_{21}}{(\Gamma_a)_{22}} \Big] (\hat{h}-\hat{h}_{\rm e}^{(1)}) \nonumber \\
&\simeq -(\Gamma_a)_{11} (\hat{h}-\hat{h}_{\rm e}^{(1)}). \label{he}
\end{align}
Hence the decay rate $\gamma_{a2}$ for $q\ll q_{\rm mc}$ in eq.~(\ref{gammaA2low}) corresponds 
to the relaxation of $\hat{h}$ to the quasi-equilibrium state $\hat{h}_{\rm e}^{(1)}$, 
while $\phi$ is frozen and $\rho$ instantly decays to $\rho^{(2)}_{\rm e}$. 
In this regime, we have $\gamma_{a2}\simeq (\Gamma_a)_{11}\simeq \sigma q/(4\eta)\sim q$ 
for $q\ll q^*$, and $\gamma_{a2}\simeq (\Gamma_a)_{11}\simeq (\kappa+kd^2\Omega_0)q^3/(4\eta )
\sim q^3$ for $q^* \ll q \ll q_{\rm mc}$. 
For $q\gg q_{\rm mc}$, on the other hand, $\gamma_{a2}$ is associated with the relaxation of 
$\rho$ towards $\rho_{\rm e}^{(1)}$, while $\phi$ is frozen and $\hat{h}$ instantly decays to
$\hat{h}_{\rm e}^{(2)}$. 
In this regime, we have $\gamma_{a2}\simeq(\Gamma_a)_{22}\simeq k(2-\Lambda_3)q^2/(4b)\sim q^2$.

From eqs.~(\ref{gammaA1low}) and (\ref{gammaA2low}), we see that the mode crossing occurs 
around $q\simeq q_{\rm mc}$; the fastest mode is associated with $\rho$ for $q< q_{\rm mc}$ 
while it is dominated by $\hat{h}$ for $q> q_{\rm mc}$.
Such a mode crossing behavior between the density and the curvature was predicted by 
Seifert and Langer for single-component lipid bilayer membranes without any surface 
tension~\cite{Seifert}. 
In Table \ref{TabRates}(a), we present a list of the approximate expressions for 
$\gamma_{a1}$ and $\gamma_{a2}$ when the membrane tension is small (the threshold 
tension $\sigma_{\rm t}$ in the table caption is defined in eq.~(\ref{sigmac}) below).

We now discuss the slowest decay rate $\gamma_{a3}$. 
Assuming $\hat{h}$ and $\rho$ vary much faster than $\phi$, we substitute  
$\hat{h}\simeq\hat{h}_{\rm e}^{(1)}$ and $\rho\simeq \rho_{\rm e}^{(1)}$ into the 
third row of eq.~(\ref{Dmatrixa}) to obtain
\begin{equation}
\frac{\partial \phi}{\partial t}\simeq -\gamma_\phi^* \phi.
\end{equation}
With the aid of eqs.~(\ref{he1}) and (\ref{rhoe1}), the effective decay rate 
$\gamma_\phi^*$ in the above equation can be obtained as
\begin{equation}
\gamma_\phi^*=\frac{L_\phi (\det A)  q^2}{2(A_{11}A_{22}-A_{12}^2)}. 
\label{gammaR}
\end{equation}
In the small and large wave number limits, its asymptotic behaviors are 
\begin{equation}
\gamma_\phi^*\to
\left\{
\begin{array}{l }
L_\phi k\tau_aq^2/2\sim q^2 \quad (q\to 0), \\
L_\phi c\Delta_\lambda q^4 \sim q^4 \quad (q\to \infty),
\end{array}
\right. 
\label{gammaPlim}
\end{equation}
where the reduced temperature $\tau_a$ is defined by~\cite{ReducedT}
\begin{equation}
\tau_a=  2\Lambda_1-\Lambda_4-\frac{(\Lambda_2-\Lambda_5)^2}{2-\Lambda_3}, 
\label{taua}
\end{equation}
and $\Delta_\lambda$ was defined before in eq.~(\ref{stab_highQ}).
When the stability conditions in eqs.~(\ref{stab1})--(\ref{stab3}) are satisfied, 
one can show that $\tau_a$ is positive. 
As the unstable region is approached, $\tau_a$ becomes smaller and eventually vanishes 
just at the boundary. 
Then the anti-registered instability in fig.~\ref{FigSchematic}(a) takes place at 
the boundary as well as in the unstable region.

The crossover wave number $q_{a}$ between the two limits in eq.~(\ref{gammaPlim}) is 
given by
\begin{equation}
q_{a}=\sqrt{\frac{k\tau_a}{2c\Delta_\lambda}}. 
\label{qac}
\end{equation}
In figs.~\ref{FigrateAlow1}(b) and \ref{FigrateAlow2}(b), we have also plotted 
$\gamma_\phi^*$. 
We see that $\gamma_\phi^*$ provides a perfect fit to the slowest mode $\gamma_{a3}$. 
Thus $\gamma_{a3}$ corresponds to the relaxation rate of $\phi$, while $\hat{h}$ and $\rho$ 
instantly change to their equilibrium values $\hat{h}_{\rm e}^{(1)}$ and $\rho_{\rm e}^{(1)}$, 
respectively.  
In Table \ref{TabRates}(c), the approximate expression for the slowest rate $\gamma_{a3}$ is 
summarized.

In fig.~\ref{FigrateAlow1}(b), we see that the bare rate $(\Gamma_a)_{33}$ almost coincides 
with the effective rate $\gamma_\phi^*\simeq \gamma_{3a}$.
This can be understood as follows. 
For the parameters used in fig.~\ref{FigrateAlow1}, the reduced temperature is approximately
given by $\tau_a\simeq 2\Lambda_1-\Lambda_4$, while we have $A_{33}\simeq k(2\Lambda_1-\Lambda_4)$ 
when the membrane is far from the unstable region (see eq.~(\ref{A33})). 
We thus have $\gamma_\phi^*\simeq L_\phi A_{33}q^2/2 =(\Gamma_a)_{33}$.
The crossover wave number given by eq.~(\ref{qac}) is 
$q_{a}=1.52\times 10^{7}$ $\mathrm{cm}^{-1}$ which is microscopic and may not be measurable in experiments.
On the other hand, the parameters used in fig.~\ref{FigrateAlow2} yield $\tau_a=3.33\times 10^{-3}$ and 
$q_{a}=3.94\times 10^6$ $\mathrm{cm}^{-1}$. 
Hence the crossover from $\gamma_{a3}\sim q^2$ to $\sim q^4$ is measurable as in usual 
near critical fluids~\cite{Onuki}. 
Note that the $q^4$-dependence in large wave numbers is not due to the coupling 
with the other modes, but is just a consequence of diffusion when there are 
squared-gradient terms in the free energy as in eq.~(\ref{Fgrad}).

%%%%%%%%%%%%%%%%%%%%%%%%%%%%%%%%%%%%%%%%%%%%%%%%%%%%%%%%%%%%%%%
\subsubsection{Eigenmodes of $\Gamma_a$: moderate tension case}

\begin{figure}[tbh]
\includegraphics[scale=0.7]{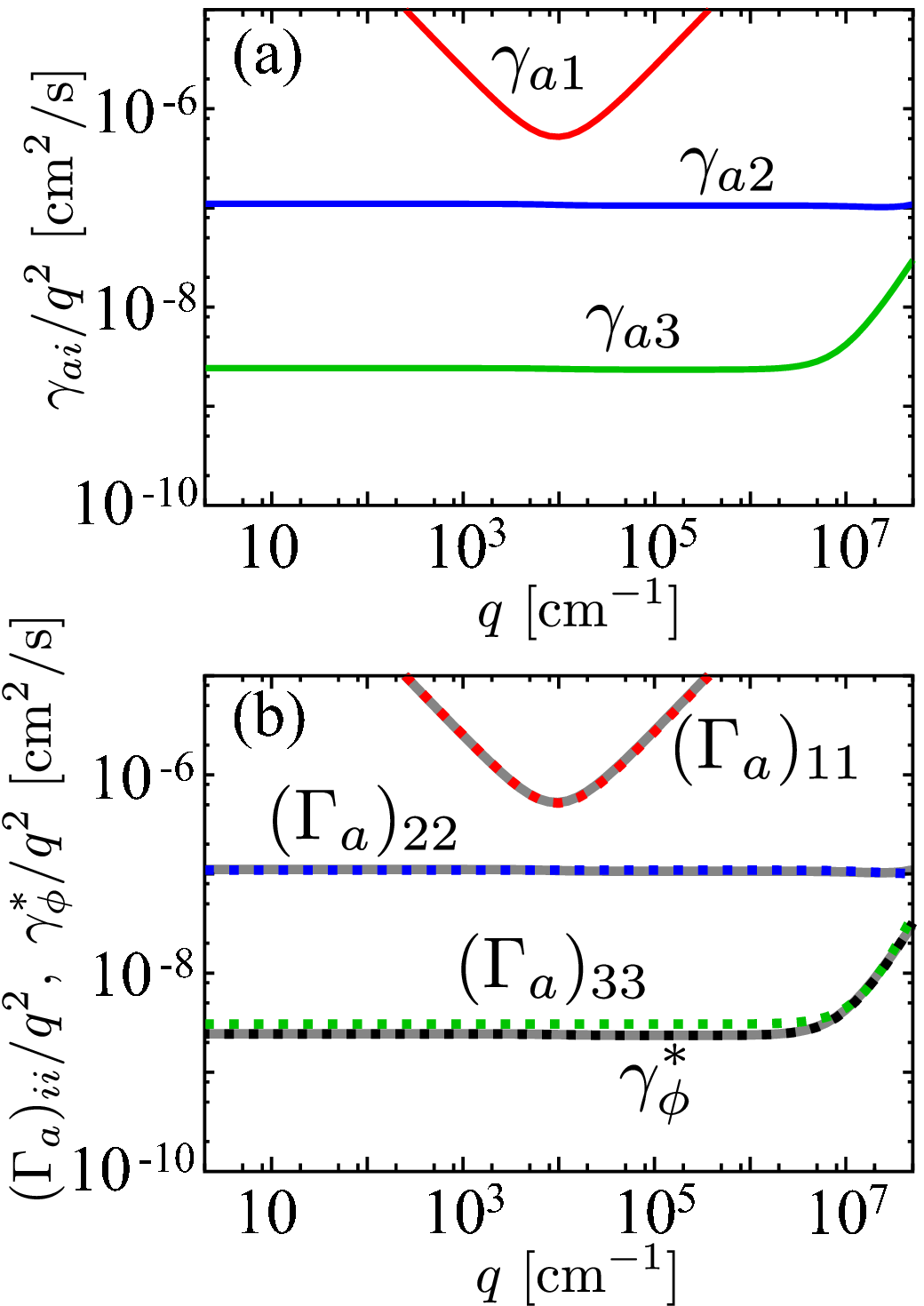}
\caption{Eigenmodes of $\Gamma_a$ for moderate tension case; 
$\sigma=10^{-4}$ $\mathrm{erg}/\mathrm{cm}^2$.
The parameter values are $(\Lambda_2, \Lambda_5)=(0.193, 0.233)$ as 
marked (A) in fig.~\ref{FigDiagram2}(d), and the membrane is not close to the 
anti-registered instability boundary.
(a) Plots of the relaxation rates $\gamma_{ai}$ as a function of the 
wave number $q$.
(b) Plots of the diagonal elements $(\Gamma_a)_{ii}$ of the matrix $\Gamma_a$ 
as a function of the wave number $q$ (dashed color lines).
The effective decay rate $\gamma_\phi^*$ is plotted with a black dashed line. 
For comparison, the relaxation rates $\gamma_{ai}$ in (a) are also plotted with 
grey solid lines.
}
\label{FigrateAmoderate1}
\end{figure}

\begin{figure}[tbh]
\includegraphics[scale=0.7]{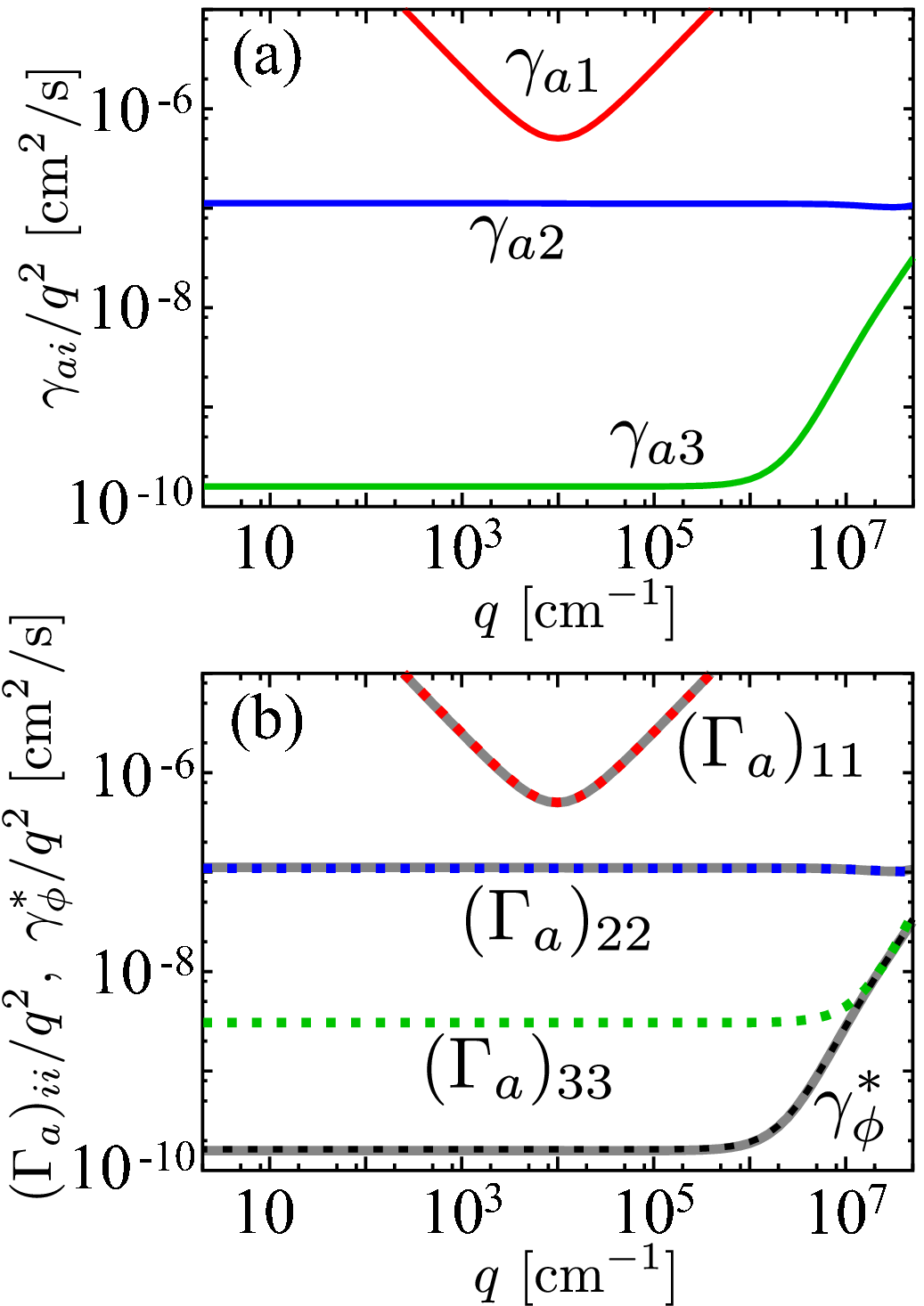}
\caption{Eigenmodes of $\Gamma_a$ for moderate tension case; 
$\sigma=10^{-4}$ $\mathrm{erg}/\mathrm{cm}^2$.
The parameter values are $(\Lambda_2, \Lambda_5)=(-0.023, 0.063)$ as 
marked (B) in fig.~\ref{FigDiagram2}(d), and the membrane is close to the 
anti-registered instability.
(a) Plots of the relaxation rates $\gamma_{ai}$ as a function of the 
wave number $q$.
(b) Plots of the diagonal elements $(\Gamma_a)_{ii}$ of the matrix $\Gamma_a$ 
as a function of the wave number $q$ (dashed color lines).
The effective decay rate $\gamma_\phi^*$ is plotted with a black dashed line. 
For comparison, the relaxation rates $\gamma_{ai}$ in (a) are also plotted with 
grey solid lines.
}
\label{FigrateAmoderate2}
\end{figure}

In figs.~\ref{FigrateAmoderate1} and \ref{FigrateAmoderate2}, we show (a) the relaxation 
rates $\gamma_{ai}$ and (b) the diagonal elements $(\Gamma_a)_{ii}$ 
of $\Gamma_a$ for a moderate surface tension, $\sigma=10^{-4}$ $\mathrm{erg}/\mathrm{cm}^2$. 
In these plots, all the parameters except $\sigma$ are the same as in figs.~\ref{FigrateAlow1} 
and \ref{FigrateAlow2}. 
In the whole wave number range, the decay rates can be approximated as 
\begin{align}
&\gamma_{a1}\simeq (\Gamma_a)_{11}, \\
&\gamma_{a2}\simeq (\Gamma_a)_{22}-\frac{(\Gamma_a)_{12}(\Gamma_a)_{21}}{(\Gamma_a)_{11}}\simeq (\Gamma_a)_{22}, \\
&\gamma_{a3}\simeq \gamma_\phi^*,
\end{align}
where $\gamma_\phi^*$ was defined in eq.~(\ref{gammaR}). 
The fastest decay rate $\gamma_{a1}$ is associated with the relaxation of $\hat{h}$ 
to $\hat{h}_{\rm e}^{(2)}$, while $\rho$ and $\phi$ are frozen. 
The second decay rate $\gamma_{a2}$ corresponds to the relaxation of 
$\rho$ to $\rho_{\rm e}^{(1)}$, while $\phi$ is frozen and $\hat{h}$ instantly changes 
to $\hat{h}_{\rm e}^{(1)}$.
The slowest decay mode $\phi$ relaxes by the effective decay rate $\gamma_\phi^*$, while 
$\hat{h}$ and $\rho$ instantly change to $\hat{h}_{\rm e}^{(1)}$ and $\rho_{\rm e}^{(1)}$, 
respectively.

The slowest decay rate $\gamma_{3a}\simeq \gamma_\phi^*$ in figs.~\ref{FigrateAmoderate1} and \ref{FigrateAmoderate2} is almost the same as in figs.~\ref{FigrateAlow1} and \ref{FigrateAlow2} 
for which the membrane tension is very small (see Table \ref{TabRates}(c)). 
However, unlike in figs.~\ref{FigrateAlow1} and \ref{FigrateAlow2}, the mode crossing behavior 
between the two fast (bending and density) modes does not occur for the moderate tension case.  
Recently, the absence of the mode crossing behavior due to the membrane tension has 
been reported in the experiment~\cite{Mell}, and theoretically discussed for 
single-component lipid bilayer membranes~\cite{JBNLM}.

Since the minimum of $(\Gamma_a)_{11}/q^2$ is located around $q\sim q^*$ and 
$(\Gamma_a)_{22}/q^2$ is almost constant, the condition $q_{\rm mc}\simeq q^*$
gives the threshold surface tension
\begin{equation}
\sigma_{\rm t} \simeq \frac{1}{\kappa} \left[ \frac{k\eta (2-\Lambda_3)}{b}\right]^2,
\label{sigmac}
\end{equation}
below which the mode crossing occurs. 
For $\Lambda_3=15/8$ and other parameter values, we can estimate 
$\sigma_{\rm t}\approx 1.91\times 10^{-5}$ $\mathrm{erg/cm}^2$. 
Table~\ref{TabRates}(a) and (b) summarize the approximate expressions of the two fastest 
rates of $\Gamma_a$ for small tension case ($\sigma<\sigma_{\rm t}$) and for large tension 
case ($\sigma>\sigma_{\rm t}$), respectively. 
Notice that in the small tension case, we always have $q^*<q_{\rm mc}$.

\begin{table}[tbh]
\caption{
Approximate expressions for the decay rates.  
(a) The two fastest decay rates $\gamma_{a1}$ and $\gamma_{a2}$ associated with $\Gamma_a$
for the small tension case $\sigma<\sigma_{\rm t}$.
(b) The two fastest decay rates $\gamma_{a1}$ and $\gamma_{a2}$ associated with $\Gamma_a$
for the moderate (larger) tension case $\sigma>\sigma_{\rm t}$.
The threshold tension $\sigma_{\rm t}$ is defined in eq.~(\ref{sigmac}). 
(c) The slowest decay rate $\gamma_{a3}$ associated with $\Gamma_a$ for both the small and the moderate tension cases.
(d) The slowest decay rate $\gamma_{b2}$ associated with $\Gamma_b$ for both the small and the moderate tension cases. 
The characteristic wave numbers $q^*$, $q_{\rm mc}$, $q_a$ and $q_b$ are defined in 
eqs.~(\ref{qstar}), (\ref{qmc}), (\ref{qac}) and (\ref{qbc}) respectively.
\label{TabRates}}
\begin{tabular}[t]{c |c c c }
\hline \hline
(a) &  \ $q\ll q^* $\quad  & \quad  $q^*\ll q\ll q_{\rm mc}$ \quad & $ q_{\rm mc}\ll q$ \quad  \\
\hline
\parbox[c][1.0cm][c]{0cm}{} $\gamma_{a1}$ & \multicolumn{2}{c}{$\displaystyle \frac{k(2-\Lambda_2)q^2}{4b}$} & $\displaystyle \frac{(\kappa+kd^2\Omega_0)q^3}{4\eta}$ \\
\parbox[c][1.0cm][c]{0cm}{} $\gamma_{a2}$ & \quad $\displaystyle \frac{\sigma q}{4\eta}$ \quad \quad & $\displaystyle \frac{(\kappa+kd^2\Omega_0)q^3}{4\eta}$  & $\displaystyle \frac{k(2-\Lambda_2)q^2}{4b}$ \\
\hline \hline
\end{tabular}

\vspace{4mm}

\begin{tabular}[t]{c |c c  }
\hline \hline
(b)  &  \ $q\ll q^* $\quad  & \quad  $q^*\ll q$ \quad  \\
\hline
\parbox[c][1.0cm][c]{0cm}{} $\gamma_{a1}$ &  $\displaystyle \frac{\sigma q}{4\eta}$ & \quad$\displaystyle \frac{(\kappa+kd^2\Omega_0)q^3}{4\eta}$ \\
\parbox[c][0.9cm][c]{0cm}{} $\gamma_{a2}$ &   \multicolumn{2}{c}{$\displaystyle \frac{k(2-\Lambda_2)q^2}{4b}$} \\
\hline \hline
\end{tabular}

\vspace{4mm}

\begin{tabular}[t]{c |c c  }
\hline \hline
(c) &  \ $q\ll q_a $\quad  & \quad  $q_a \ll q$ \quad  \\
\hline
\parbox[c][0.8cm][c]{0cm}{} $\gamma_{a3}$ &  $\frac{1}{2}L_\phi k\tau_aq^2 $ & \quad$L_\phi c\Delta_\lambda q^4$ \\
\hline \hline
\end{tabular}

\vspace{4mm}

\begin{tabular}[t]{c |c c  }
\hline \hline
(d) &  \ $q\ll q_b $\quad  & \quad  $q_b \ll q$ \quad  \\
\hline
\parbox[c][0.8cm][c]{0cm}{} $\gamma_{b2}$ &  $\frac{1}{2}L_\phi k\tau_b q^2 $ & \quad$L_\phi c\Delta_\lambda q^4$ \\
\hline \hline
\end{tabular}
\end{table}

%%%%%%%%%%%%%%%%%%%%%%%%%%%%%%%%%%%%%%%%
\subsubsection{Eigenmodes of $\Gamma_b$}

\begin{figure}[tbh]
\includegraphics[scale=0.7]{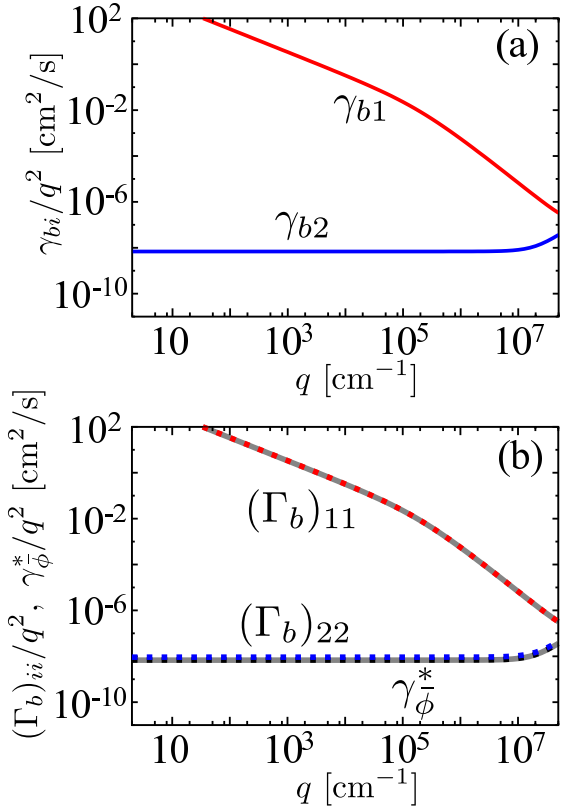}
\caption{
Eigenmodes of $\Gamma_b$ when
the parameter values are $(\Lambda_2, \Lambda_5)=(0.193, 0.233)$ as 
marked (A) in fig.~\ref{FigDiagram2}(d), and the membrane is not close to the 
anti-registered instability boundary.
(a) Plots of the relaxation rates $\gamma_{bi}$ ($i=1,2$) as a function of the 
wave number $q$.
(b) Plots of the diagonal elements $(\Gamma_b)_{ii}$ ($i=1,2$) of the matrix $\Gamma_b$ 
as a function of the wave number $q$ (dashed color lines).
The effective decay rate $\gamma_{\bar\phi}^*$ is plotted with a black dashed line. 
For comparison, the relaxation rates $\gamma_{bi}$ in (a) are also plotted with 
grey solid lines.
}
\label{FigrateB1}
\end{figure}

\begin{figure}[tbh]
\includegraphics[scale=0.7]{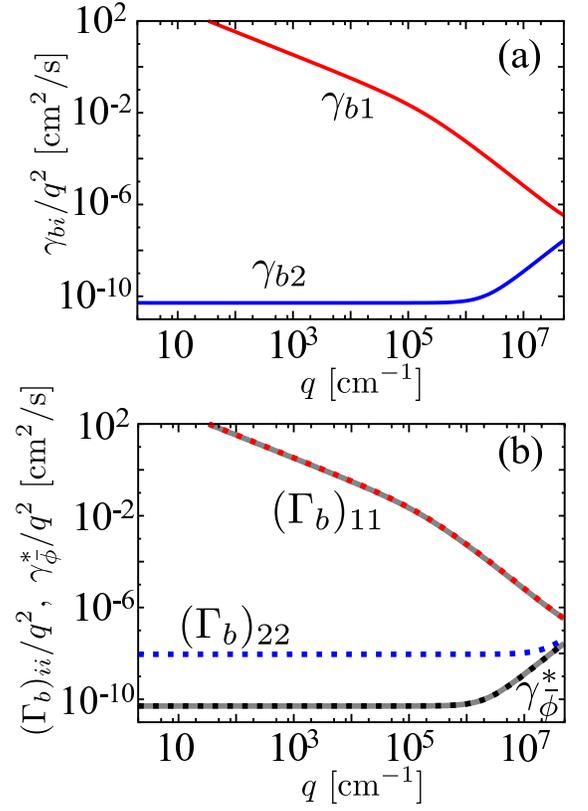}
\caption{
Eigenmodes of $\Gamma_b$ when 
the parameter values are $(\Lambda_2, \Lambda_5)=(0.445, 0.405)$ as 
marked (C) in fig.~\ref{FigDiagram2}(d), and the membrane is close to the registered 
instability boundary.
(a) Plots of the relaxation rates $\gamma_{bi}$ as a function of the 
wave number $q$.
(b) Plots of the diagonal elements $(\Gamma_b)_{ii}$ of the matrix $\Gamma_b$ 
as a function of the wave number $q$ (dashed color lines).
The effective decay rate $\gamma_{\bar\phi}^*$ is plotted with a black dashed line. 
For comparison, the relaxation rates $\gamma_{bi}$ in (a) are also plotted with 
grey solid lines.
}
\label{FigrateB2}
\end{figure}

In figs.~\ref{FigrateB1} and \ref{FigrateB2}, we plot the eigenvalues and diagonal elements 
of $\Gamma_b$ in eq.~(\ref{gamma_b}). 
The parameters are chosen as $(\Lambda_2, \Lambda_5)=(0.193, 0.233)$ in fig.~\ref{FigrateB1}
and $(0.445, 0.405)$ in fig.~\ref{FigrateB2}.  
These two choices are marked with (A) and (C) in fig.~\ref{FigDiagram2}(d).
Since $\Gamma_b$ is a $2\times 2$ matrix, its eigenvalues $\gamma_b$ 
can be easily obtained as
\begin{align}
\gamma_b=&\frac{1}{2} \Big[ (\Gamma_b)_{11}+(\Gamma_b)_{22} \nonumber \\
& \pm \sqrt{\{ (\Gamma_b)_{11}-(\Gamma_b)_{22}\}^2+4(\Gamma_b)_{12}(\Gamma_b)_{21}} \Big] \\
\simeq & \frac{1}{2} \left[ \Tr \Gamma_b \pm \{ \Tr\Gamma_b +2(\Gamma_b)_{11}^{-1} \det\Gamma_b\}\right],
\end{align}
where the second equality follows from $(\Gamma_b)_{11}\gg (\Gamma_b)_{22}$ and 
$(\Gamma_b)_{11}^2 \gg (\Gamma_b)_{12}(\Gamma_b)_{21}$. 
Then we obtain approximately
\begin{align}
&\gamma_{b1}\simeq (\Gamma_b)_{11}, \\
&\gamma_{b2}\simeq \frac{L_\phi q^2 \det B}{2B_{11}}\equiv \gamma_{\bar\phi}^*. 
\label{gammaB2}
\end{align}

Equating the right hand side of eq.~(\ref{Dmatrixb}) to zero, we obtain the quasi-equilibrium variables as 
\begin{align}
&\bar\rho_{\rm e} (\bar\phi,q)=-\frac{B_{12}}{B_{11}}\bar\phi, \label{rhobare}\\
&\bar\phi_{\rm e} (\bar\rho,q)=-\frac{B_{21}}{B_{22}}\bar\rho. \label{phibare}
\end{align}
As in the previous subsections, the fastest decay rate 
$\gamma_{b1}\simeq (\Gamma_b)_{11}$ is associated with the relaxation of $\bar\rho$ to 
$\bar\rho_{\rm e}$ while $\bar\phi$ is frozen. 
However, as shown in figs.~\ref{FigrateB1} and \ref{FigrateB2}, the decay rate $\gamma_{b1}$ 
is very large, and our theory, in which inertial effect is neglected, may not properly 
describe the dynamics of such a small time scale~\cite{Seifert}.
Hence we do not further discuss the wave number dependence of $\gamma_{b1}$.

Nevertheless, we can discuss the slower relaxation of $\bar\phi$ because $\bar\rho$ relaxes 
rapidly to the quasi-equilibrium value $\bar\rho_{\rm e}(\bar\phi)$.
With the aid of eq.~(\ref{rhobare}), substitution of $\bar\rho\simeq \bar\rho_{\rm e}(\bar\phi)$ 
into the second row of eq.~(\ref{Dmatrixb}) yields 
$\partial \bar\phi /\partial t\simeq -\gamma_{\bar\phi}^* \bar\phi$. 
From eq.~(\ref{gammaB2}), we see that the slower decay rate 
$\gamma_{b2}\simeq \gamma_{\bar\phi}^*$ corresponds to the relaxation of $\bar\phi$, 
while $\bar\rho$ instantly changes to $\bar\rho_{\rm e}(\bar\phi,q)$.
In the small and large wave number limits, the asymptotic behaviors are 
\begin{equation}
\gamma_{\bar\phi}^*\to
\left\{
\begin{array}{l }
L_\phi k\tau_bq^2/2 \sim q^2\quad \quad (q\to 0), \\
L_\phi c\Delta_\lambda q^4 \sim q^4\quad (q\to \infty),
\end{array}
\right. 
\label{gammaPBlim}
\end{equation}
where the other reduced temperature $\tau_b$ is defined by~\cite{ReducedT}
\begin{equation}
\tau_b=  2\Lambda_1+\Lambda_4-\frac{(\Lambda_2+\Lambda_5)^2}{2+\Lambda_3}. 
\label{taub}
\end{equation}
When the stability conditions in eqs.~(\ref{stab1}), (\ref{stab2}) and (\ref{stab4}) are
satisfied, $\tau_b$ is positive in the stable region.
As the unstable region is approached, $\tau_b$ becomes smaller and eventually vanishes 
at the boundary where the registered instability in fig.~\ref{FigSchematic}(b) starts 
to take place.
The crossover wave number $q_{b}$ between the two limits in eq.~(\ref{gammaPBlim}) is given by
\begin{equation}
q_{b}=\sqrt{\frac{k\tau_b}{2c\Delta_\lambda}}. 
\label{qbc}
\end{equation}
When the system is away from the unstable region ($\tau_b=0.141$) as in fig.~\ref{FigrateB1}, 
we have $q_{b}=2.57\times 10^7$ $\mathrm{cm}^{-1}$ which is too large to be observed. 
However, when the system is close to the unstable region ($\tau_b=1.10\times 10^{-3}$) as 
in fig.~\ref{FigrateB2}, we have $q_{b}=2.27\times 10^6$ $\mathrm{cm}^{-1}$ which is 
measurable in experiments. 
In Table \ref{TabRates}(d), the approximate expressions for the slowest rate $\gamma_{b2}$ 
are summarized.


%%%%%%%%%%%%%%%%%%%%%%%%%%%%%%%%%%%%%%%
\subsection{Domain relaxation dynamics}

\begin{figure}[tbh]
\includegraphics[scale=0.75]{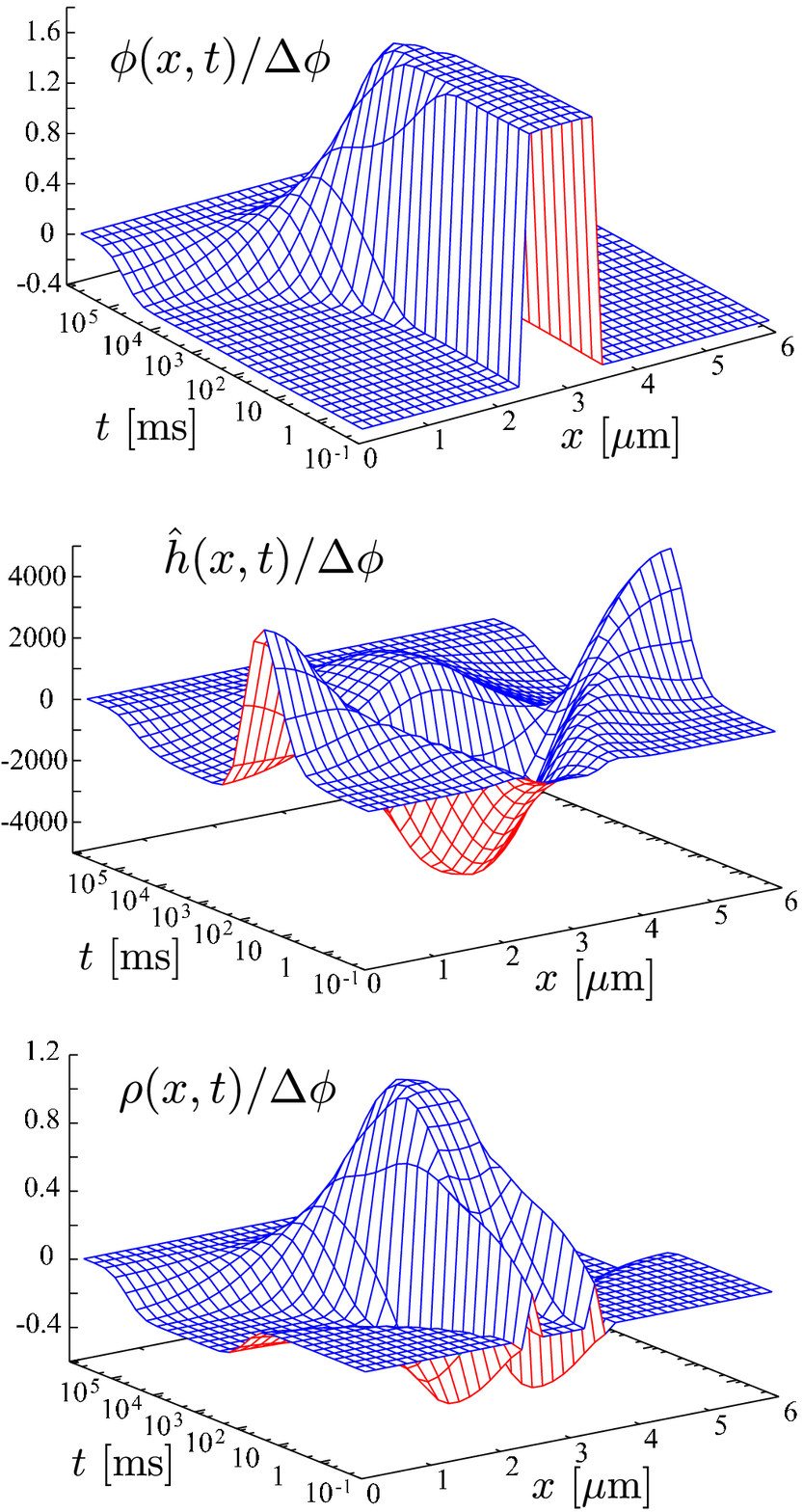}
\caption{Time evolutions of $\phi(x,t)/\Delta \phi$, $\hat{h}(x)/\Delta \phi$ and 
$\rho(x,t)\Delta \phi$ from the initial state given by eqs.~(\ref{initialR1}) and (\ref{initialR2}). 
The parameters are the same as in fig.~\ref{FigrateAlow2}, {\it i.e.}, 
$\sigma=10^{-8}$ $\mathrm{erg/cm}^{-2}$ and $(\Lambda_2, \Lambda_5)=(-0.023, 0.063)$.
}
\label{FigTimeR}
\end{figure}

In this subsection, we examine the relaxation dynamics of a domain in which $\phi$ is 
larger than the outside. 
When the system is in the stable region as in eqs.~(\ref{stab1}) and (\ref{stab2}), such 
a domain should relax to a homogeneous state $(\hat{h}, \rho, \phi) =0$. 
Let us assume that the initial state $(\hat{h}_0,\rho_0,\phi_0)$ at $t=0$ is described by 
one-dimensional profiles 
\begin{align}
&\hat{h}_0(x)=\rho_0(x)=0,
\label{initialR2} \\
&\phi_0(x)=\frac{\Delta\phi}{2} \left[\phi_{\rm c}+ \tanh \left\{
\frac{ L_{\rm d}-|2x-L|}{2\ell} \right\} \right],
\label{initialR1} 
\end{align}
while these profiles are homogeneous in $y$-direction.
The profile $\phi_0(x)$ represents a patch centered at $x=L/2$, and its size and 
interfacial thickness are given by $L_{\rm d}$ and $\ell$, respectively. 
The difference of $\phi$ between the inside and the outside the initial domain 
is given by $\Delta\phi$, whereas $\phi_{\rm c}$ is determined so that the spacial 
average of $\phi_0$ vanishes.
We will not discuss the other variables $(\bar\rho,\bar\phi)$ because they are not 
coupled to $(\hat{h},\rho,\phi)$.

The three variables can be generally expressed as Fourier series defined by 
\begin{equation}
g (x,t)=\sum_{n=-\infty}^{\infty} \exp \left( \frac{2\pi i n x}{L} \right) g(q_n,t),
\end{equation}
where 
\begin{equation}
q_n=\frac{2\pi n}{L}.
\end{equation}
Let ${\bm a}_0(q_n)= (0, 0, \phi_0(q_n))$ denote the Fourier modes of the initial state 
given by eqs.~(\ref{initialR2}) and (\ref{initialR1}). 
Since the time evolution of each Fourier mode is governed by eq.~(\ref{Dmatrixa}), we 
can write ${\bm a}(q_n, t)=e^{-\Gamma_a(q)t}\,{\bm a}_0(q_n)$ with $q=|q_n|$.
The matrix $\Gamma_a(q)$ can be diagonalized by using its eigenvalues $\gamma_i(q)$ 
and their respective eigenvectors ${\bm w}_i(q)$ as
\begin{equation}
\Lambda_a(q)=W^{-1}(q) \Gamma_a(q) W(q),
\end{equation} 
where $\Lambda_a = {\rm diag}(\gamma_{a1}, \gamma_{a2}, \gamma_{a3})$ 
is the diagonalized matrix and $W = ({\bm w}_1,{\bm w}_2,{\bm w}_3)$. 
Then ${\bm a}(x,t)$ can be generally written as
\begin{equation}
{\bm a}(x, t)=\sum_{n=-n_{\rm c}}^{n_{\rm c}} 
\exp\left( {\frac{2\pi i n x}{L}} \right) W e^{-\Lambda_a(q)t}W^{-1}{\bm a}_0(q_n), 
\label{TimeSol}
\end{equation}
where we have introduced a cut-off wave number set by the monolayer thickness $d$
\begin{equation}
\frac{2\pi n_{\rm c} }{L}=\frac{\pi}{d}.
\end{equation}

In fig.~\ref{FigTimeR}, we present the time evolution of $\phi(x,t)$, $\hat{h}(x,t)$ and 
$\rho(x,t)$ obtained from eq.~(\ref{TimeSol}) by setting $L=6000$, $\ell=10$ and 
$L_{\rm d}=1000$ in nm.
The other parameters are the same as in fig.~\ref{FigrateAlow2} and the system is close 
to the anti-registered instability. 
Notice that ${\bm a}(x,t)$ divided by $\Delta\phi$ is independent of $\Delta\phi$ 
since eq.~(\ref{Dmatrixa}) is linear in ${\bm a}(x,t)$. 
For $t\le 10^{3}$ ms, $|\hat{h}|$ and $|\rho|$ increase while $\phi$ remains almost the same. 
This means that, within a small time interval, non-zero $\phi$ induces the bending 
$\hat{h}$ and the density difference $\rho$ which were initially both zero. 
For $t \ge 10^3$ ms, all the three variables become smaller and almost vanish 
for $t \ge 10^5$ ms.

\begin{figure}
\includegraphics[scale=0.7]{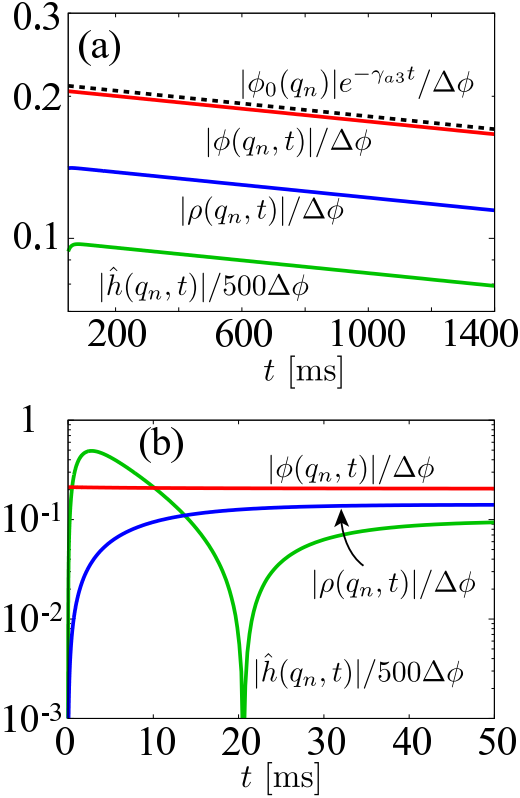}
\caption{Semilogarithmic plots of the time evolutions of 
$|\hat{h}(q_n,t)|/\Delta \phi$, 
$|\rho(q_n,t)|/\Delta \phi$ and 
$|\phi(q_n,t)|/\Delta \phi$ 
for (a) large $t$ and (b) small $t$. 
Here we choose $q_n=3.14$ $\mu\mathrm{m}^{-1}$.
In (a), we also plot $|\phi_0(q_n)|e^{-\gamma_{a3}t}$ with a dashed line.}
\label{FigTimeMode1}
\end{figure}

\begin{figure}
\includegraphics[scale=0.7]{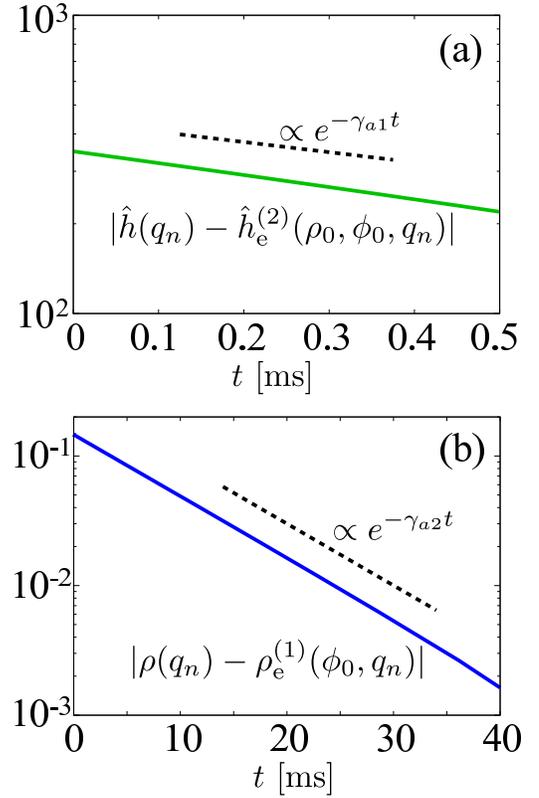}
\caption{Semilogarithmic plots of the time evolutions of
(a) $|\hat{h}(q_n,t)-\hat{h}_{\rm e}^{(2)}(\rho_0,\phi_0,q_n)|$ (solid line) and 
$e^{-\gamma_{a1}t}$ (dashed line), and 
(b) $|\rho(q_n,t)-\rho_{\rm e}^{(1)}(\phi_0,q_n)|$ (solid line) and 
$e^{-\gamma_{a2}t}$ (dashed line).
}
\label{FigTimeMode2}
\end{figure}

The above dynamics can be roughly understood by looking at the time evolution of a 
Fourier mode at $q_n \simeq 2\pi /L_{\rm d}$. 
In figs.~\ref{FigTimeMode1}(a) and (b), the time evolutions of $|\hat{h}(q_n)|$, 
$|\rho(q_n)|$ and $|\phi(q_n)|$ are presented at $q_n= 3.14$ $\mu\mathrm{m}^{-1}$ for which the 
decay rates are $\gamma_{a1}=785$ $\mathrm{s}^{-1}$, 
$\gamma_{a2}=110$ $\mathrm{s}^{-1}$ and $\gamma_{a3}=0.157$ $\mathrm{s}^{-1}$. 
In fig.~\ref{FigTimeMode2}, $|\rho -\rho_{\rm e}^{(1)}(\phi_0)|$ and 
$|\hat{h}-\hat{h}_{\rm e}^{(2)}(\rho_0,\phi_0)|$ are plotted as a function of $t$ for the 
same wave number as in fig.~\ref{FigTimeMode1}.
As for the long time behavior, $t\gg \gamma_{a2}^{-1}$ ($\gg\gamma_{a1}^{-1}$), 
fig.~\ref{FigTimeMode1}(a) shows that all the three variables decay exponentially with a 
common decay rate $\gamma_{a3}$. 
In this regime, we have $\hat{h}(q_n)\simeq\hat{h}_{\rm e}^{(1)}(\phi, q_n)$, 
$\rho(q_n)\simeq\rho_{\rm e}^{(1)}(\phi, q_n)$ and
\begin{align}
\phi(q_n)\simeq \phi_0(q_n)e^{-\gamma_{a3}t}\simeq \phi_0(q_n)e^{-\gamma_\phi^*t}, 
\label{phiTapp}
\end{align}
as in the the discussion after eq.~(\ref{qac}). 
Substituting eq.~(\ref{phiTapp}) into eqs.~(\ref{he1}) and (\ref{rhoe1}), we then have
\begin{align}
&\hat{h}(q_n) \simeq\frac{ A_{12} A_{23}- A_{13} A_{22}}{ A_{11}A_{22}- A_{12}^2} 
\phi_0e^{-\gamma_{a3}t}, \\
&\rho(q_n) \simeq\frac{ A_{12} A_{13}- A_{11} A_{23}}{ A_{11}A_{22}- A_{12}^2} \phi_0e^{-\gamma_{a3}t},
\end{align}
which decay exponentially with the common rate $\gamma_{a3}$.

For shorter times, on the other hand, $|\hat{h}|$ and $|\rho|$ rapidly vary while $\phi$ 
stays almost constant, as shown in fig.~\ref{FigTimeMode1}(b). 
In figs.~\ref{FigTimeMode1} and \ref{FigTimeMode2}, the chosen $q_n=3.14$ $\mu\mathrm{m}^{-1}$ 
is much larger than the mode crossing wave number $q_{\rm mc}=0.438$ $\mu\mathrm{m}^{-1}$. 
Then the fastest decay rate $\gamma_{a1}$ corresponds to the 
relaxation of $\hat{h}$ to $\hat{h}_{\rm e}^{(2)}$ while $\rho$ and $\phi$ are frozen. Notice that in fig.~\ref{FigTimeMode1}(b) $\hat{h}(q_n,t)$ changes its sign around $t\approx 20\ [\mathrm{ms}]$.
Hence $\hat{h}-\hat{h}_{\rm e}^{(2)}(\rho_0,\phi_0)$ decays exponentially with the rate 
$\gamma_{a1}$ for $t\ll \gamma_{a2}^{-1}$ ($\ll \gamma_{a3}^{-1}$).
However, fig.~\ref{FigTimeMode2}(a) shows a slight deviation between 
$\hat{h}-\hat{h}_{\rm e}^{(2)}(\rho_0,\phi_0)$ and $e^{-\gamma_{a1}t}$. 
This is due to the fact that the ratio $\gamma_{a1}/\gamma_{a2}=7.15$ is not large enough 
to regard $\rho$ as a completely frozen variable. 
The second mode $\gamma_{a2}$ in fig.~\ref{FigTimeMode2}(b) is associated with the 
relaxation of $\rho$ to $\rho_{\rm e}^{(1)}$, while $\phi$ is frozen and $\hat{h}$ rapidly 
changes to $\hat{h}_{\rm e}^{(2)}$. 
Hence we have $\rho-\rho_{\rm e}^{(1)}(\phi_0) \sim e^{-\gamma_{a2}t}$ for 
$\gamma_{a3}^{-1} \ll t \ll \gamma_{a1}^{-1}$.

%%%%%%%%%%%%%%%%%%%%%%%%%%%%%%%%
\section{Summary and Discussion}
%%%%%%%%%%%%%%%%%%%%%%%%%%%%%%%%
\label{summary}

In this paper, 
we have theoretically investigated the relaxation dynamics of a binary lipid bilayer 
membrane by taking into account (i) the coupling between the height and the density 
variables, (ii) the hydrodynamics of the surrounding fluid, (iii) the frictional force 
between the upper and lower leaflets, and (iv) the mutual diffusion in each monolayer. 
In sect.~\ref{freeenegy}, we have constructed the free energy in terms of the membrane 
shape $h$, the total lipid density $\rho^\pm$, and the lipid density difference $\phi^\pm$ 
up to quadratic order. 
The membrane surface tension $\sigma$, which was neglected in the previous theory
for single-component lipid bilayer membranes~\cite{Seifert}, and taken into account recently~\cite{JBNLM}, naturally appears in the expansion of the 
general free energy in eq.~(\ref{Ftot}).

In sect.~\ref{dynamics}, the dynamic equations have been formulated on the basis of 
momentum and molecular number conservations. 
In Appendix \ref{appa}, we have proved the non-negative definiteness of the dissipation
in our formulation. 
We have also presented an alternative derivation of the dynamic equations by using the 
Onsager's variational principle in Appendix \ref{appb}. 
The derived equations for binary lipid bilayer membranes are the generalization of 
those in the Seifert and Langer model~\cite{Seifert}. 
We have further obtained the relaxation equations for five variables by integrating 
out the velocity field of the surrounding fluid (see also Appendix C).
The equations are separated into two independent sets of equations; one for 
$(\hat{h}, \rho, \phi)$ and the other for $(\bar\phi,\bar\rho)$. 
The former equations change their signs under the interchange of the upper and lower 
leaflets, while the latter equations are invariant.

In sect.~\ref{results},
we have discussed the stability of the one phase state and found that 
there are two possible instabilities; the anti-registered instability of
$(\hat{h}, \rho, \phi)$ and the registered instability of $(\bar\phi,\bar\rho)$.
We have investigated in detail the relaxation rates of the various hydrodynamic modes.
In the case of small surface tension $\sigma<\sigma_{\rm t}$ (see Eq~(\ref{sigmac})),
figs.~\ref{FigrateAlow1} and \ref{FigrateAlow2} show that the mode crossing between 
$\rho$ and $\hat{h}$ takes place around the intermediate wave number $q_{\rm mc}$.
Such a mode crossing was originally predicted for tensionless single-component lipid 
bilayers~\cite{Seifert}.
When $\sigma>\sigma_{\rm t}$, however, the height variable $\hat{h}$ is the fastest 
mode in the whole wave number range, and the mode crossing does not occur~\cite{JBNLM}.

Unlike single-component membranes for which either $\hat{h}$ or $\rho$ is the slowest mode,  
mutual diffusion in two-component membranes is the slowest mode both for small and moderate 
surface tensions.  
While $\phi$ varies slowly, the faster variables $\hat{h}$ and $\rho$ rapidly 
approach their respective quasi-equilibrium states determined by $\phi$. 
In all the examined cases, the effective decay rate $\gamma_{a3}\simeq\gamma_\phi^*$ for 
$\phi$ (see eq.~(\ref{gammaR})) is smaller than the bare decay rate $(\Gamma_a)_{33}$ 
because of the faster slaved variables $\hat{h}$ and $\rho$.

As the unstable region is approached, the slowdown of the effective rate $\gamma_{a3}$ becomes even more 
significant, and the crossover from $\gamma_{a3}\sim q^2$ to $\sim q^4$ 
behaviors may be measurable in experiments. 
As for the faster dynamics, the relaxation of $\hat{h}$ is controlled by the hydrodynamics 
of the surrounding fluid, and the corresponding decay rate is approximately given by 
$A_{11}/(4\eta d^2 q)$ (see eqs.~(\ref{gamma_a}) and (\ref{he})). 
The relaxation of $\rho$ is dominated by the inter-monolayer friction, and its decay 
rate is given by $A_{22}q^2/(4b)$ (see the sentences after eq.~(\ref{he})).
We have also examined the relaxation of a domain that is rich in $\phi$ when the 
membrane is close to the unstable region. 
In the very early stage, the bending of the membrane is induced by a non-zero density 
variation of $\phi$ even the membrane is initially flat. 
In the late stage of the relaxation process, all the variables decay with the common decay
rate $\gamma_{a3}\simeq\gamma_\phi^*$ as mentioned above.

The dynamics of $(\bar\rho,\bar\phi)$ is simpler than that of $(\hat{h}, \rho, \phi)$. 
The fastest variable $\bar\rho$ instantly approaches to its quasi-equilibrium state 
$\bar\rho_{\rm e}$. 
Then $\bar\phi$ relaxes with the effective decay rate $\gamma_{b2}\simeq \gamma_{\bar\phi}^*$
(see eq.~(\ref{gammaB2})) which becomes even slower as the unstable region is approached.

While the kinetic parameters and some of the static parameters have been determined in Sec.~\ref{results}, the dimensionless parameters $\Lambda_i$ ($i=1,3,4,5$) in the free energy could not be estimated from the previous experimental data. However, the behaviors of the relaxation rates, which are summarized in Table \ref{TabRates} and are described in figs.~\ref{FigrateAlow1}--\ref{FigrateB2}, are not sensitive to these parameters, unless the reduced temperatures $\tau_a$ and $\tau_b$ are very close to zero ($\tau_a$ or $\tau_b$ are defined in terms of $\Lambda_i$'s in eqs.~(\ref{taua}) and (\ref{taub}), respectively). In fact, besides the parameters determined from the experimental data, these reduced temperatures are the only relevant parameters. In the case of $\Gamma_a$ (resp.~$\Gamma_b$), this is because the time scales of the different modes characterized by the diagonal elements $(\Gamma_a)_{ii}$ (resp.~$(\Gamma_b)_{ii}$) are well separated, except in the vicinity of the characteristic wave number $q\simeq q_{\rm mc}$ where the values of two fastest modes of $\Gamma_a$ become close in the low tension case ($q_{\rm mc}$ is independent of the parameters that could not be estimated). 
The two reduced temperatures measure the distances in the phase space from their respective critical points \cite{ReducedT}, and one can experimentally control them by varying the average lipid composition and the temperature. As discussed above, when $\tau_a$ (resp.~$\tau_b$) is close to zero, the anti-registered (resp.~registered) diffusive mode becomes very slow, and the associated rate is given by $\gamma^*_\phi$ (resp.~$\gamma^*_{\bar\phi}$).

Finally, we give some remarks.
(i) We have constructed our free energy as a power series expansion up to quadratic  
order with respect to the deformation and the densities about the reference state. 
Here the physical meaning and microscopic interpretation of some phenomenological 
coupling parameters such as $\Lambda_i$ are not so obvious.
It would be ideal to construct a free energy from a microscopic model, and perform 
a series expansion of the free energy with respect to the densities and curvature.
With such a procedure, a connection between our phenomenological parameters and the 
microscopic quantities can be made. 
Recently, an attempt has been made for a flat bilayer membrane by Williamson and 
Olmsted who derived a mean field free energy from a semi-microscopic lattice bilayer 
model.
In their model, the difference in length between the two different lipid species 
was taken into account~\cite{Olmsted}.

(ii) In real biological cells, inclusions in membranes such as proteins play essential roles. 
It was recently discussed that the proteins which span the bilayer give rise to a further 
constraint in the dynamics and an additional source of dissipation leading to anomalous 
diffusion~\cite{JBnew}. 
Furthermore, the surrounding fluid can be viscoelastic rather than purely viscous, and  
inclusions can be active in a sense that they consume energy and drive membranes out of 
equilibrium. 
Neglecting the bilayer structure, some authors have investigated the membrane shape 
fluctuations when it contains active/non-active inclusions and is surrounded by 
viscoelastic media~\cite{Granek,Lau,KomuraJPCM}. 
Generalization of our theory to such situations is also interesting.

(iii) As we further approach the unstable region or the critical point, the dynamical 
non-linear coupling (mode-mode coupling) between the density variables and the velocity 
fields in the bilayer becomes important like in the ordinary 3D critical 
fluids~\cite{SKI07,Inaura,RKSI11,KellerDynamics}.  
It would be interesting to investigate the effects of the bilayer structure and friction
on top of the mode coupling between the velocity and the density fields.

%%%%%%%%%%%%%%%%%%%%%%% 
\begin{acknowledgments}
%%%%%%%%%%%%%%%%%%%%%%%

We thank D.\ Andelman, T.\ Hoshino, T.\ Kato, C.-Y. D. Lu, P.\ D.\ Olmsted, P.\ Sens, 
M. Turner, K. Yasuda for useful discussions.
R.O. and S.K. acknowledge support from the Grant-in-Aid for Scientific Research on
Innovative Areas ``\textit{Fluctuation and Structure}" (Grant No.\ 25103010) from the Ministry
of Education, Culture, Sports, Science, and Technology of Japan,
the Grant-in-Aid for Scientific Research (C) (Grant No.\ 24540439)
from the Japan Society for the Promotion of Science (JSPS),
and the JSPS Core-to-Core Program ``\textit{International Research Network
for Non-equilibrium Dynamics of Soft Matter}".
\end{acknowledgments}

\appendix
%%%%%%%%%%%%%%%%%%%%%%%%%%%%%%
\section{Dissipation function}
%%%%%%%%%%%%%%%%%%%%%%%%%%%%%%
\label{appa}

In this Appendix, we discuss the dissipation which is related to 
the change rate of the free energy, $W=-{\rm d}F/{\rm d}t$.
The contribution due to the mutual diffusion is given by the change of $\phi^\pm$ 
\begin{equation}
(\dot F)_{\phi^\pm}\equiv \int {\rm d}^2x \, \frac{\delta F}{\delta \phi^\pm}
\frac{\partial\phi^\pm}{\partial t}=-\int {\rm d}^2x \, \frac{| {\bm j}^\pm_\phi |^2}{L_\phi}, 
\label{dFphi}
\end{equation}
where use has been made of eqs.~(\ref{phicon}) and (\ref{Dflux}).

Next we examine the contribution from the change of $\rho^\pm$. 
Using eqs.~(\ref{rhocon}) and (\ref{lateralforce}), we obtain 
\begin{align}
(\dot F)_{\rho^\pm}\equiv & \int {\rm d}^2x \, \frac{\delta F}{\delta \rho^\pm}
\frac{\partial\rho^\pm}{\partial t}
=-\int{\rm d}^2x\,{\bm f}^\pm\cdot\tilde{\bm v}^\pm
\nonumber \\
=&\int {\rm d}^2x \, \left[ -\tilde{\cal D}^\pm_{\rm v} \pm \tilde{\bm v}^\pm\cdot 
\left\{ \tensor{T}^\pm\cdot {\bm e}_z - b(\tilde {\bm v}^+-\tilde {\bm v}^-)\right\}\right],  
\label{dFrho1}
\end{align}
where ${\bm e}_z$ is the unit vector in the $z$-direction, and $\tilde{\cal D}^\pm_{\rm v}$ is 
the viscous dissipation in the monolayers
\begin{align}
\tilde{\cal D}^\pm_{\rm v} =& \sum_{ij}(\partial_j \tilde v_i^\pm) \tau^\pm_{ij} 
\nonumber \\
=& \sum_{ij} \frac{\mu}{2}\left( \partial_i \tilde v_j^\pm+
\partial_j \tilde v_i^\pm-\delta_{ij}\tilde\nabla\cdot \tilde{\bm v}^\pm \right)^2 
+\zeta \left(\tilde\nabla\cdot \tilde{\bm v}^\pm \right)^2. 
\label{dissipationLayer}
\end{align}

From the boundary conditions eqs.~(\ref{BC1}) and (\ref{BC2}), the velocity in the monolayers 
$\tilde{\bm v}^\pm$ can be expressed in terms of the surrounding fluid velocity ${\bm v}^\pm$ as  
$\tilde{\bm v}^\pm={\bm v}^\pm-(\partial h/\partial t){\bm e}_z$. 
Using this relation with eqs.~(\ref{Stokes}) and (\ref{incompressible}), we obtain
\begin{equation}
\int {\rm d}^2x \,  \tilde{\bm v}^\pm\cdot \tensor{T}^\pm \cdot {\bm e}_z 
= \mp \int_\pm {\rm d}^3x \ {\cal D}_{\rm v} -\int {\rm d}^2x \, 
\frac{\partial h}{\partial t} T^\pm_{zz}, 
\label{dFrho2}
\end{equation}
where $\int_\pm {\rm d}^3x$ denotes the 3D integration in the ranges of $z>0$ and $z<0$, 
and ${\cal D}_{\rm v}$ is the viscous dissipation in the surrounding fluid
\begin{equation}
{\cal D}_{\rm v}= \sum_{ij} \frac{\eta}{2}\left( \partial _i v_j+\partial _j  v_i \right)^2. \label{dissipationOut}
\end{equation}
Furthermore, we define the dissipation due to the friction between the two monolayers
\begin{equation}
\tilde{\cal D}_{\rm f}= b|\tilde{\bm v}^+-\tilde{\bm v}^- |^2.
\end{equation}

Combining eqs.~(\ref{dFphi}), (\ref{dFrho1}) and (\ref{dFrho2}), we finally obtain
\begin{align}
W=& -\int {\rm d}^2x \, \frac{\delta F}{\delta h}
\frac{\partial h}{\partial t} -\sum_{\epsilon=+,-} 
\left[ (\dot F)_{\rho^\epsilon}+ (\dot F)_{\phi^\epsilon} \right] \nonumber \\
=& \int {\rm d}^2x \, \Big[  \tilde{\cal D}_{\rm f}+\sum_{\epsilon=+,-}\big\{ \tilde{\cal D}^\epsilon_{\rm v} +L_\phi^{-1} | {\bm j}_\phi^\epsilon|^2 \big\} \Big] \nonumber \\
& + \sum_{\epsilon=+,-} \int_\epsilon {\rm d}^3x \, {\cal D}_{\rm v}. 
\label{dissipationTot}
\end{align}
Here we see that the dissipation occurs through (i) the friction 
between the monolayers, (ii) the shear and bulk viscosity of the monolayers, (iii) the mutual 
diffusion in the monolayers, and (iv) the shear viscosity of the surrounding fluid. 
The positivity of $\eta$, $\mu$, $\zeta$, $L_\phi$ and $b$ ensures the positivity of the 
dissipation; $W\ge 0$.

%%%%%%%%%%%%%%%%%%%%%%%%%%%%%%%%%%%%%%%%%
\section{Onsager's variational principle}
%%%%%%%%%%%%%%%%%%%%%%%%%%%%%%%%%%%%%%%%%
\label{appb}

In Appendix \ref{appa}, we have derived the dissipation eq.~(\ref{dissipationTot}), 
starting from the dynamic equations given by eqs.~(\ref{Stokes}), (\ref{incompressible}), 
(\ref{rhocon}), (\ref{phicon}), (\ref{Dflux}), (\ref{lateralforce}), (\ref{normalforce}), 
(\ref{BC1}) and (\ref{BC2}). 
Conversely, these dynamic equations can be obtained by the variational principle provided 
that we know the dissipations, namely, the right hand side of 
eq.~(\ref{dissipationTot}). 
This is called the Onsager's variational principle~\cite{Onsager1,Onsager2,Doi}. 
It is applicable to many dynamical problems in soft matter 
such as  colloidal dispersions, membranes and polymer solutions if inertial effects 
can be neglected~\cite{Doi,JBNLM,JBnew,Arroyo,Rahimi}.

More precisely, we can derive eqs.~(\ref{Stokes}), (\ref{Dflux}), (\ref{lateralforce}) 
and (\ref{normalforce}) by minimizing the Rayleighian
\begin{align}
{\cal R}=& \frac{W}{2}+\frac{{\rm d}F}{{\rm d}t} \nonumber \\
=& \frac{W}{2}+\int  {\rm d}^2x \left[ \frac{\delta F}{\delta h}\dot h+\sum_{\epsilon=+,-} 
\left\{  \frac{\delta F}{\delta \rho^\epsilon}\dot\rho^\epsilon+ 
\frac{\delta F}{\delta \phi^\epsilon}\dot\phi^\epsilon \right\}\right]
\end{align}
with respect to ${\bm v}$, $\tilde{\bm v}^\pm$, ${\bm j}_\phi^\pm$, 
$\dot h= \partial h/\partial t$, $\dot\rho^\pm= \partial \rho^{\pm}/\partial t$ 
and $\dot\phi^\pm= \partial \phi^{\pm}/\partial t$.

The incompressible condition of the surrounding fluid (eq.~(\ref{incompressible})), the 
continuity equations (eqs.~(\ref{rhocon}) and (\ref{phicon})), and the non-slip boundary 
conditions (eqs.~(\ref{BC1}) and (\ref{BC2})) are taken into account as the constraints 
under which the Rayleighian is minimized. 
Hence we minimize the shifted Rayleighian
\begin{align}
{\cal R}^*= &\ {\cal R}-\sum_{\epsilon=+,-}\int_\epsilon {\rm d}^3x\  p\nabla\cdot{\bm v} 
\nonumber \\
&+\int {\rm d}^2x\sum_{\epsilon=+,-} \left[ C_\rho^\epsilon 
(\dot\rho^\epsilon+\tilde\nabla\cdot\tilde{\bm v}^\epsilon )+
C_\phi^\epsilon (\dot\phi^\epsilon+\tilde\nabla\cdot{\bm j}_\phi^\epsilon )\right]
\nonumber \\
&+\int {\rm d}^2x\sum_{\epsilon=+,-} \left[ C_{vz}^\epsilon (\dot h-v_z^\epsilon)+
\sum _{i=x,y} C_{vi}^\epsilon (\tilde v_i^\epsilon- v_i^\epsilon )\right]
\end{align}
with respect to ${\bm v}$, $\tilde{\bm v}^\pm$, ${\bm j}_\phi^\pm$, $\dot h$, $\dot\rho^\pm$, 
$\dot\phi^\pm$ and the Lagrange multipliers $p$, $C_\rho^\pm$, $C_\phi^\pm$, $C_{vi}^\pm$ 
($i=x,y,z$).

We first consider an infinitesimal variation of ${\bm v}$ as ${\bm v}\mapsto {\bm v}+\delta {\bm v}$. 
Then the first variation $\delta {\cal R}^*|_{{\bm v}}$ with respect to ${\bm v}$ is given by
\begin{align}
\delta {\cal R}^*|_{{\bm v}}=&-\sum_{\epsilon=+,-}\int_\epsilon  {\rm d}^3x \, 
\left[ -\nabla p+\nabla^2 {\bm v}+\nabla (\nabla\cdot {\bm v}) \right]\cdot \delta{\bm v}  
\nonumber \\
& -\int  {\rm d}^2x \, \sum_{\epsilon=+,-} \sum_{i=x,y,z} \left[ C_{vi}^\epsilon+\epsilon T^\epsilon_{iz}\right]  \delta v_i^\epsilon,
\end{align}
where $\delta v_i^\pm$ is the velocity variation $\delta v_i$ evaluated at $z\to \pm 0$. 
Hence the minimization of ${\cal R}^*$ with respect to ${\bm v}$ (in the bulk region) and 
$p$ yields the Stokes equation for the surrounding fluid, eq.~(\ref{Stokes}).
Here the Lagrange multiplier $p$ can be identified as the pressure field.

Furthermore, minimizing ${\cal R}^*$ with respect to ${\bm v}$ at  $z=\pm 0$, we obtain
\begin{equation}
C_{vi}^\pm=\mp T_{iz}^\pm.
\label{v_min}
\end{equation}
Similarly, minimization of  ${\cal R}^*$ with respect to 
$\tilde{\bm v}^\pm$, ${\bm j}_\phi^\pm$, $\dot h$, $\dot\rho^\pm$ and $\dot\phi^\pm$ yields
\begin{align}
&C_{vi}^\pm-\partial_j\tau_{ij}^\pm-\partial_i C_\rho^\pm  \pm b(\tilde v_i^+-\tilde v_i^-) =0, 
\label{vm_min}\\
& {\bm j}_\phi^\pm=L_\phi\nabla C_\phi^\pm, 
\label{j_min}\\
& C_{vz}^+ +C_{vz}^-=-\frac{\delta F}{\delta h}, 
\label{h_min}\\
& C_\rho^\pm=-\frac{\delta F}{\delta \rho^\pm}, 
\label{rho_min}\\
& C_\phi^\pm=-\frac{\delta F}{\delta \phi^\pm}, 
\label{phi_min}
\end{align}
respectively. 
Substituting eqs.~(\ref{v_min}) and (\ref{rho_min}) into eq.~(\ref{vm_min}), we obtain the 
force balance equation of the upper and lower monolayers in the tangential direction, 
eq.~(\ref{lateralforce}). 
The force balance equation in the normal direction, eq.~(\ref{normalforce}), is obtained by 
substituting eq.~(\ref{v_min}) into eq.~(\ref{h_min}). 
Finally, substitution of eq.~(\ref{phi_min}) into eq.~(\ref{j_min}) yields the diffusive 
flux, eq.~(\ref{Dflux}).

%%%%%%%%%%%%%%%%%%%%%%%%%%%%%%%%%%%%%%%%%%%
\section{Elimination of the velocity field}
%%%%%%%%%%%%%%%%%%%%%%%%%%%%%%%%%%%%%%%%%%%
\label{appc}

In this appendix, we discuss the dynamics of a single Fourier mode. 
Without loss of generality, we can take the $(x,y)$-coordinate so that the direction 
of the wave vector $\tilde{\bm q}$ coincides with the $x$-direction, 
{\it i.e.}, $\tilde{\bm q}= (q_x, 0)$ with $q_x>0$. 
Then we have $q=|\tilde{\bm q}|=q_x$.
Substitution of $p(x,z) =p(q,z) e ^{iqx}$ and ${\bm v}(x,z) ={\bm v} (q,z) e ^{iqx}$
into eq.~(\ref{Stokes}) and (\ref{incompressible}) gives
\begin{align}
&iqp+\eta(q^2-\partial_z^2) v_x=\partial_zp+\eta (q^2-\partial_z^2) v_z=0, \\
&(q^2-\partial_z^2)v_y=0,\\
&\partial_z v_z+iqv_x=0.
\end{align}
We then solve these equations to have
\begin{align}
&p=2\eta R^\pm e^{\mp qz},  \label{sol1} \\
&v_z=(R^\pm z+ S) e^{\mp qz}, \label{sol2} \\
&v_x=\frac{i}{q} [R^\pm \mp q(R^\pm z+S)] e^{\mp qz}, \label{sol3}  \\
&v_y=Q^\pm e^{\mp qz}.\label{sol4}
\end{align}
where $Q^\pm$, $R^\pm$ and $S$ are integral constants, and the upper and the 
lower signs indicate the fluids for $z>0$ and $z<0$, respectively. 
In deriving eqs.~(\ref{sol2}) and (\ref{sol3}), we have used the boundary condition 
$v_z^+=v_z^-$ (Eq. (\ref{BC2})).

Next we substitute $\rho^\pm(x) =\rho^\pm(q) e ^{iqx}$, $\phi^\pm(x) =\phi^\pm(q) e ^{iqx}$, 
$h(x) =h(q) e ^{iqx}$ and $\tilde {\bm v}^\pm(x) =\tilde {\bm v}^\pm(q) e ^{iqx}$ into
eqs.~(\ref{lateralforce}) and use eqs.~(\ref{BC1}) and (\ref{sol1})--(\ref{sol4}). 
After some algebra we obtain
\begin{align}
&R^+-R^--2qS=-2iq c_1{\cal F}_{\tilde{\bm q}}[ f_x^+-f_x^-], \label{Cint1}\\
&R^++R^-=-2iq c_2{\cal F}_{\tilde{\bm q}}[ f_x^++f_x^-], \label{Cint2} \\
&Q^\pm=0,
\end{align}
where ${\cal F}_{\tilde{\bm q}}[\cdots]$ denotes the Fourier transform at wave number 
$\tilde{\bm q}=(q,0)$, and $c_1$ and $c_2$ are defined in eqs.~(\ref{c1}) and (\ref{c2}), respectively.
Similarly, from eqs.~(\ref{normalforce}), (\ref{sol1}) and (\ref{sol2}), we obtain 
$S=-{\cal F}_{\tilde{\bm q}} [\delta F /\delta h]/(4\eta q)$. 
Furthermore we use eqs.~(\ref{BC2}) and (\ref{sol2}) to have $\partial h(q)/\partial t =v_z(q,0)=S$. 
Then the time evolution of $h(q)$ is given by
\begin{equation}
\frac{\partial h}{\partial t}=-\frac{1}{4\eta q}{\cal F}_{\tilde{\bm q}}\left[\frac{\delta F}{\delta h}\right]. 
\label{h_rate}
\end{equation}

Substituting $\rho^\pm(x) =\rho^\pm(q) e ^{iq x}$ and 
$\tilde {\bm v}^\pm(x) =\tilde {\bm v}^\pm(q) e ^{iq x}$ into eq.~(\ref{rhocon}) and 
using eqs.~(\ref{BC1}) and (\ref{sol3}), we have 
$\partial \rho^\pm(q)/\partial t=-iq \tilde v_x^\pm=R^\pm\mp q S$. 
Then $\rho(q)$ and $\bar\rho (q)$ defined in eq.~(\ref{mrho}) obey the following equations
\begin{align}
&\frac{\partial \rho}{\partial t}=-iq c_1{\cal F}_{\tilde{\bm q}}[f_x^+-f_x^-], \label{rho_rate}\\
&\frac{\partial \bar\rho}{\partial t}=-iq c_2 {\cal F}_{\tilde{\bm q}}[ f_x^++f_x^-], \label{rhobar_rate}
\end{align}
where use has been made of eqs.~(\ref{Cint1}) and (\ref{Cint2}). 
The time evolution of $\phi^\pm(q)$ in eq.~(\ref{phicon}) now reads
\begin{equation}
\frac{\partial \phi^\pm}{\partial t} = -L_\phi q^2 {\cal F}_{\tilde{\bm q}}
\left[ \frac{\delta F}{\delta \phi^\pm}\right].
\label{phi_rate}
\end{equation}

Functional derivatives of eqs.~(\ref{Fcoup}) and (\ref{Fgrad}) are given by 
\begin{align}
\frac{\delta F}{\delta \rho^\pm}=&\ \frac{k}{2}\left(2\alpha^\pm +\Lambda_2\beta^\pm+\Lambda_3\alpha^\mp+\Lambda_5\beta^\mp\right), \nonumber\\
&-\frac{c}{2}\left(2\tilde\nabla^2\rho^\pm+\lambda_2\tilde\nabla^2\phi^\pm\right), 
\label{FDrho}
\end{align}
and
\begin{align}
\frac{\delta F}{\delta \phi^\pm}=&\ \frac{k}{2}\left(2\Lambda_1 \beta^\pm +\Lambda_2\alpha^\pm+\Lambda_4\beta^\mp+\Lambda_5\alpha^\mp\right) \nonumber\\
&-\frac{c}{2}\left(2\lambda_1\tilde\nabla^2\phi^\pm+\lambda_2\tilde\nabla^2\rho^\pm\right), 
\label{FDphi}
\end{align}
respectively.
Using eqs.~(\ref{flayer}), (\ref{normalforce}), (\ref{FDrho}) and (\ref{FDphi}), we can 
calculate the right hand sides of eqs.~(\ref{h_rate})--(\ref{phi_rate}) to obtain 
eqs.~(\ref{Dmatrixa})--(\ref{gamma_b}).

%%%%%%%%%%%%%%%%%%%%%%%%%%%%%%%%%
\section{Thermodynamic stability}
%%%%%%%%%%%%%%%%%%%%%%%%%%%%%%%%%
\label{appd}

%%%%%%%%%%%%%%%%%%%%%%%%%%%%%%%%%%%%%%%%%%%%%%%%%%%%
\subsection{Stability at $q\to 0$ and $q\to \infty$}

The static fluctuations and stability of the system is characterized by the eigenvalues 
of the matrices $A(q)$ and $B(q)$ in eq.~(\ref{FFourier}). 
We define the susceptibilities $\chi_a(q)$ and $\chi_b(q)$ as the reciprocals of the 
eigenvalues of $A(q)$ and $B(q)$, respectively.
For large wave numbers, they behave as
\begin{align}
\frac{1}{\chi_a(q)}= q^4 \times
\left\{
\begin{array}{l}
(\kappa +kd^2\Omega_0 )d^2 +\sigma d^2 q^{-2} +O(q^{-4}), \\
cM^\pm q^{-2}+O(q^{-4}).
\end{array}
\right.
\end{align}
and 
\begin{align}
\frac{1}{\chi_b(q)}= q^2 [cM^\pm +O(q^{-2})],
\end{align}
where
\begin{equation}
M^\pm =1+\lambda_1 \pm \sqrt{(1-\lambda_1)^2+\lambda_2^2}.
\end{equation}
The thermodynamic stability in large wave numbers is ensured by $M^\pm>0$, 
which is equivalent to eq.~(\ref{stab_highQ}).

For small wave numbers, the eigenvalues of $A$ are given by
\begin{align}
\frac{1}{\chi_a(q)}= 
\left\{
\begin{array}{l }
\sigma d^2 q^2 +\kappa d^2 q^4+O(q^6), \\
(1/\chi_a^\pm )+O(q^2),
\end{array}
\right. \label{chi_a}
\end{align}
where $\chi_a^\pm$ are given by
\begin{align}
\frac{1}{\chi_a^\pm}=\frac{1}{2} 
\left[ A_{22}+A_{33}\pm \sqrt{(A_{22}-A_{33})^2+4A_{23}^2} \right], 
\end{align}
and $A_{ij}$ are evaluated at $q=0$. 
In eq.~(\ref{chi_a}), the first line vanishes as $q\to 0$. 
This zero eigenvalue at $q=0$ corresponds to the homogeneous translation of the membrane 
in the $z$-direction, which costs no energy.

The eigenvalues of $B$ are given by
\begin{align}
\frac{1}{\chi_b(q)}=  \frac{1}{\chi_b^\pm}+O(q^2), \label{chi_b}
\end{align}
where
\begin{align}
&\frac{1}{\chi_b^\pm}=\frac{1}{2} \left[ B_{11}+B_{22}\pm 
\sqrt{(B_{11}-B_{22})^2+4B_{12}^2} \right],
\end{align}
where $B_{ij}$ are evaluated at $q=0$. 
The thermodynamic stability conditions at $q=0$ are given by $\chi^-_a$, $\chi^-_b>0$ 
(note that $\chi^+_a$, $\chi^+_b>0$ if $\chi^-_a$, $\chi^-_b>0$). 
These inequalities are ensured if and only if eqs.~(\ref{stab1})--(\ref{stab4}) are satisfied.

%%%%%%%%%%%%%%%%%%%%%%%%%%%%%%%%%%%%%%%%%%%%%%%%%%%
\subsection{Stability at intermediate wave numbers}

Assuming that the modes at $q=0$ and $\infty$ are thermodynamically stable, we have 
$A_{22}(q)$, $B_{11}(q)>0$ from eqs.~(\ref{stab_highQ}) and (\ref{stab1}). 
Then, $\hat{h}$, $\rho$ and $\bar\rho$ can be integrated out from the Boltzmann weight 
$\exp(-F[\hat{h},\rho,\phi,\bar\rho,\bar\phi ]/k_{\rm B}T)$ to obtain an effective free energy 
$F_{\rm eff}[\phi,\bar\phi] = -k_{\rm B}T\ln [ \int {\cal D}\hat{h}\,{\cal D}\rho \,{\cal D}\bar\rho \, 
e^{-F/k_{\rm B}T}]$, where $\int {\cal D}\hat{h}\,{\cal D}\rho \, {\cal D}\bar\rho$ 
denotes the functional integral and $k_{\rm B}$ is the Boltzmann constant.
Since our free energy $F$ is quadratic, $F_{\rm eff}$ can be obtained by minimizing $F$ with 
respect to $\hat{h}$, $\rho$ and $\bar\rho$. 
By equating the right hand sides of eqs.~(\ref{normalforce}) and (\ref{FDrho}) to zero, 
we can eliminate $\hat{h}$, $\rho$ and $\bar\rho$ from eq.~(\ref{FFourier}) to obtain
\begin{equation}
F_{\rm eff}=\int \frac{{\rm d}^2q}{(2\pi)^2}\frac{1}{2}\Bigg[\frac{|\phi|^2}{\chi_\phi(q)}+\frac{|\bar\phi|^2}{\chi_{\bar\phi}(q)}\Bigg].
\end{equation}
Here the susceptibilities $\chi_\phi(q)$ and $\chi_{\bar\phi}(q)$ for $\phi$ and $\bar\phi$ 
are given by
\begin{align}
\frac{1}{\chi_\phi(q)}=\frac{\det A}{A_{11}A_{22}-A_{12}^2},
\end{align}
and
\begin{equation}
\frac{1}{\chi_{\bar\phi}(q)}=\frac{\det B}{B_{11}},
\end{equation}
respectively.

We expand $1/\chi_\phi$ and $1/\chi_{\bar\phi}$ in powers of $q$ to have
\begin{align}
\frac{1}{k\chi_\phi(q)}=\tau_a+\left[ s_1\left( \Delta_a^2+\Delta_\lambda\right)-s_2(\nu\tau_a)^2\right] (qd)^2+O(q^4), 
\label{Sphi}
\end{align}
and
\begin{align}
\frac{1}{k\chi_{\bar\phi}(q)}=\tau_b +s_1\left( \Delta_b^2+\Delta_\lambda\right) (qd)^2+O(q^4), \label{Sphibar}
\end{align}
where the reduced temperatures $\tau_a$ and $\tau_b$ were defined in 
eqs.~(\ref{taua}) and (\ref{taub}), respectively. 
In the above, we have defined dimensionless combinations
\begin{align}
&s_1= \frac{2c}{kd^2}, \quad s_2=\frac{k}{\sigma},\label{nonD1}\\
&\Delta_a= \frac{\Lambda_2-\Lambda_5}{2-\Lambda_3}-\frac{\lambda_2}{2},\\
&\Delta_b= \frac{\Lambda_2+\Lambda_5}{2+\Lambda_3}-\frac{\lambda_2}{2}.
\end{align}
Notice that the stability conditions at $q=0$ in eqs.~(\ref{stab1})-(\ref{stab4}) are 
equivalent to the conditions $\tau_a, \tau_b, A_{22}(0), B_{11}(0) >0$.

In eq.~(\ref{Sphibar}), the term quadratic in $q$ is always positive, which indicates 
that $\bar\phi$ and $\bar\rho$ do not exhibit any instability at intermediate wave 
numbers when they are stable at $q=0$. 
In eq.~(\ref{Sphi}), however, the quadratic term is negative if
\begin{equation}
\tau_a>\left[\frac{s_1(\Delta_a^2+\Delta_\lambda)}{s_2\nu^2}\right]^{1/2}\equiv \tau_a^*. 
\label{nonD2}
\end{equation} 
In this case, we need to add a quartic term $C_\phi(qd)^4$ to eq.~(\ref{Sphi}) in 
order to examine the stability at intermediate wave numbers.  
The coefficient of the quartic term is obtained as
\begin{equation}
C_\phi\simeq s_2 (s_3+s_2\nu^2\tau_a)(\nu\tau_a)^2, \label{quartic}
\end{equation}
where we have defined 
\begin{equation}
s_3 = \frac{\kappa}{\sigma d^2}.
\label{nonD3}
\end{equation}
In deriving eq.~(\ref{quartic}), we have assumed $s_1\ll s_2$, $s_3$. 
In Table \ref{TabDless}, we list the values of $s_1$, $s_2$, $s_3$ and $\tau_a^*$ for 
the parameter values chosen in sect.~\ref{results} and 
$\Delta_a=\Delta_\lambda=\nu=1$.

\begin{table}[tbh]
\caption{Values of dimensionless quantities in eqs.~(\ref{nonD1}), (\ref{nonD2}) and (\ref{nonD3}) 
evaluated for the parameter values in sect.~\ref{results} and 
$\Delta_a=\Delta_\lambda=\nu=1$.
\label{TabDless}}
\begin{ruledtabular}
\begin{tabular}[t]{c|c c c c}
$\sigma$ $(\mathrm{erg/cm}^2)$ & $s_1$ & $s_2 $ & $s_3 $ & $\tau_a^*$  \\
\hline
$10^{-4}$ & $2.86\times 10^{-2}$ & $7\times 10^5$ & $10^6$  & $2.86 \times 10^{-4}$ \\
$10^{-8}$ & $2.86\times 10^{-2}$ & $7\times 10^9 $ & $10^{10} $ & $2.86 \times 10^{-6}$ \\
\end{tabular}
\end{ruledtabular}
\end{table}

\begin{figure}[tbh]
\includegraphics[scale=0.7]{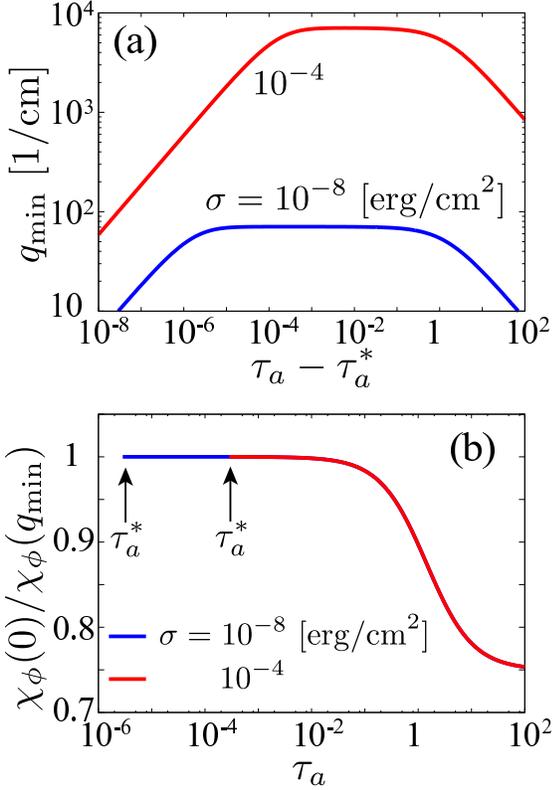}
\caption{
Plots of (a) $q_{\rm min}$ defined by eq.~(\ref{qmin}) as a function of $\tau_a-\tau_a^*$,
and (b) $\chi_\phi(0)/\chi_\phi(q_{\rm min})$ as a function of $\tau_a$.
The parameters are the same as in Table \ref{TabDless}. 
The inverse of $\chi_\phi$ exhibits a minima at $q=q_{\rm min}(>0)$ for $\tau_a>\tau_a^*$. 
We see in (b) that the minimum value $1/\chi_\phi(q_{\rm min})$ does not become zero as far 
as $1/\chi_\phi(0)=k\tau_a>0$.
The two curves in (b) for different $\sigma$ values coincide almost completely for 
$\tau_a > \tau_a^* = 2.86 \times 10^{-4}$.
}
\label{FigInter}
\end{figure}

For $\tau_a>\tau_a^*$, the reciprocal of the susceptibility $1/\chi_\phi$ has a minima 
at an intermediate wave number $q=q_{\rm min}$,
\begin{equation}
\frac{1}{k\chi_\phi(q_{\rm min})}= \tau_a-\frac{s_2^2\nu^4 [\tau_a^2-(\tau_a^*)^2]^2}{4C_\phi}, \label{Sphimin}
\end{equation}
where
\begin{equation}
q_{\rm min}= \frac{1}{d}\left[\frac{s_2\nu^2\{ \tau_a^2-(\tau_a^*)^2\}}{2C_\phi}\right]^{1/2}.
\label{qmin}
\end{equation}
From Table~\ref{TabDless}, we can assume $\tau_a^*\sim \sqrt{s_1/s_2} \ll 1$ and 
$s_2/ s_3=kd^2/\kappa$ to be of order of unity so that 
\begin{align}
\frac{1}{k\chi_\phi(q_{\rm min})}\simeq
\left\{
         \begin{array}{l }
          \displaystyle \tau_a \quad (\tau_a- \tau_a^* \ll s_3/s_2),\\
         \vspace{-3mm} \\
           \displaystyle \frac{\tau_a}{4} \left[ 3+\frac{1}{1+(s_2/s_3)\nu^2\tau_a}\right] \quad (\tau_a\gg \tau_a^*).
         \end{array}
    \right.  \label{Sphiminapp}
\end{align}
and
\begin{align}
q_{\rm min}\simeq \frac{1}{d} \times
\left\{
         \begin{array}{l }
          \displaystyle \left[ \frac{\tau_a-\tau_a^*}{(s_3+s_2\nu^2\tau_a^*)\tau_a^*}\right]^{1/2} \quad (\tau_a- \tau_a^* \ll \tau_a^*),\\
          \vspace{-2mm}\\
           \displaystyle [2(s_3+s_2\nu^2\tau_a)]^{-1/2} \quad (\tau_a\gg \tau_a^*).
         \end{array}
    \right.  \label{qminapp}
\end{align}
Since we see $1/\chi_\phi(q_{\rm min})>0$ in eq.~(\ref{Sphiminapp}), the instability at 
$q=q_{\rm min}$ does not takes place in both regimes.

In fig.~\ref{FigInter}(a), we plot $q_{\rm min}$ as a function of $\tau_a-\tau_a^*$.
For both $\sigma=10^{-8}$ and $10^{-4}$ $\mathrm{erg/cm}^2$, 
we have 
$q_{\rm min}\sim (\tau_a-\tau_a^*)^{1/2}$ for $\tau_a- \tau_a^* \ll \tau_a^*$, 
$q_{\rm min}\simeq 1/(d\sqrt{2s_3})$ for $\tau_a^* \ll \tau_a\ll s_3/(s_2\nu^2)$, and 
$q_{\rm min}\sim \tau_a^{-1/2}$ for $\tau_a \gg s_3/(s_2\nu^2)$. 
These behaviors are in good agreement with eq.~(\ref{qminapp}). 
In fig.~\ref{FigInter}(b), we plot $1/\chi_a(q_{\rm min})$ multiplied by $\chi_a(0)=1/(k\tau_a)$ 
as a function of $\tau_a$, where the parameters are the same as in (a). 
We see that the curves for different $\sigma$ values almost coincide and in agreement 
with eq.~(\ref{Sphiminapp}).
The quantity $\chi_\phi(0)/\chi_\phi(q_{\rm min})$ monotonically decreases from unity to the lower 
bound $3/4$ as $\tau_a$ increases. 
Therefore we conclude that instability does not occur at intermediate wave numbers when 
the modes are stable at $q=0$. 
Therefore the overall thermodynamic stability is ensured by the conditions in 
eqs.~(\ref{stab1})--(\ref{stab4}).

Leibler and Andelman discussed the instability in two-component membranes at 
intermediate wave numbers~\cite{Leibler}. 
Although they did not explicitly take into account the bilayer structure, their model 
is similar to ours because the composition-bending coupling is explicitly taken into account.
However, they treated the coefficients of the power series of the susceptibility as 
independent parameters.
In our study, the coefficient in eq.~(\ref{Sphi}) and eq.~(\ref{quartic}) vary simultaneously 
when $\Lambda_i$ values are changed.

%\newpage
%\bibliography{lipid}% Produces the bibliography via BibTeX.

\begin{thebibliography}{99}
\bibitem{Pressure}
Let $f(\rho_1,\cdots, \rho_k)$ denote the free energy density of a $k$-component fluid mixture 
as a function of the mass densities $\rho_i$ $(i=1,\cdots, k)$ of each component. 
Then the pressure of this mixture is given by $p=\sum_i \rho_i(\partial f/\partial \rho_i)-f$.
\bibitem{coupling}
Notice that we have already taken into account the term $(\tilde\nabla^2h)^2$ in $F_{\rm def}$.
\bibitem{ReducedT}
Because the parameters $\Lambda_i$ depend on the temperature, one may change the temperature 
in order to enter the unstable region at some temperature $T_{\rm c}$. 
Since the combination $\tau_a$ ($\tau_b$) changes sign at the transition, it is proportional 
to $T-T_{\rm c}$. 
Therefore, being dimensionless, it is a reduced temperature. 
See also eqs.~(\ref{Sphi}) and (\ref{Sphibar}) in Appendix \ref{appd} where the (normalized) 
susceptibilities for $\phi$ and $\bar\phi$ at $q=0$ are given by $\tau_a^{-1}$ and 
$\tau_b^{-1}$, respectively.
\end{thebibliography}
%%%%%%%%%%%%%%%%%%%%%%%%%%%

\end{document}